\documentclass[useAMS,usenatbib]{mn2e}
\usepackage[utf8]{inputenc}
\usepackage[english]{babel}
\usepackage{graphicx}
\usepackage{rotating}
\usepackage[section] {placeins}
\usepackage{amsmath, amssymb}
\usepackage{gensymb}
\usepackage{color}
\usepackage{multirow}
\usepackage{makeidx}
\usepackage{xspace}
\usepackage{float}
\usepackage{verbatim}

\newcommand{\apj}{ApJ}       
\newcommand{\apjl}{ApJ}          
\newcommand{\apjs}{ApJS}
\newcommand{\mnras}{MNRAS}

\newcommand{\araa}{ARA\&A}
\newcommand{\aj}{AJ}

\newcommand{\pasp}{PASP}

\newcommand{\refsec}[1]{Section~\ref{#1}}
\newcommand{\reffig}[1]{Fig.~\ref{#1}}

\newcommand{\reftab}[1]{Table \ref{#1}}

\def\Ab{Abell\xspace}
\def\atl{ATLAS$^{\rm 3D}$\xspace}

\def\Mr{$\rm{M_r}$\xspace}

\def\TS{T-$\Sigma$\xspace}
\def\epsR{$\epsilon$-$R$\xspace}

\def\logR{$\log \, R$\xspace}
\def\eps{$\epsilon$\xspace}
\def\distr{$n(\epsilon)$\xspace}
\def\sig3{$\Sigma_{3}$\xspace}
\def\logsig3{$\log \, \Sigma_3$\xspace}

\title[Galaxy ellipticity in clusters]{On the distribution of galaxy ellipticity in clusters}

\author[F.~D'Eugenio et al.]
{F. D'Eugenio$^{1,2,3}$\thanks{E-mail: francesco.deugenio@anu.edu.au},
R. C. W. Houghton$^1$,
R. L. Davies$^1$
and E. Dalla Bont\`a$^{3,4}$\\
$^1$Sub-department of Astrophysics, Department of Physics, University of
Oxford, Denys Wilkinson Building, Keble Road, Oxford OX1 3RH\\
$^2$Research School of Astronomy and Astrophysics, Australian National
University, Canberra ACT 2611, Australia\\
$^3$ARC Centre of Excellence for All-Sky Astrophysics (CAASTRO)\\
$^4$Dipartimento di Fisica e Astronomia ``G. Galilei'', Universit\`a degli
Studi di Padova, Vicolo dell'Osservatorio 3, I-35122, Padova, Italy\\
$^5$INAF Osservatorio Astronomico di Padova, Vicolo dell'Osservatorio 5,
I-35122, Padova, Italy
}

\begin{document}

    \date{\today}

    \pagerange{\pageref{firstpage}--\pageref{lastpage}} \pubyear{2012}

    \maketitle

    \label{firstpage}

  \begin{abstract}

    We study the distribution of projected ellipticity $n(\epsilon)$ for galaxies in a sample
    of 20 rich (Richness $\geq 2$) nearby ($z < 0.1$) clusters of galaxies. We find no evidence of
    differences in $n(\epsilon)$, although the nearest cluster in the sample (the Coma Cluster)
    is the largest outlier ($P(\mathrm{same}) < 0.05$). We then study $n(\epsilon)$ within the clusters, and
    find that $\epsilon$ increases with projected cluster-centric radius R (hereafter the $\epsilon$-R
    relation). This trend is preserved at fixed magnitude, showing that this relation exists
    over and above the trend of more luminous galaxies to be both rounder and more
    common in the centres of clusters. The $\epsilon$-R relation is particularly strong in the
    subsample of intrinsically flattened galaxies ($\epsilon > 0.4$), therefore it is not a consequence of
    the increasing fraction of round slow rotator galaxies near cluster centers.
    Furthermore, the $\epsilon$-R relation persists for just smooth flattened galaxies and for
    galaxies with de Vaucouleurs-like light profiles, suggesting that the variation of the
    spiral fraction with radius is not the underlying cause of the trend. We interpret our
    findings in light of the classification of early type galaxies (ETGs) as fast and slow
    rotators. We conclude that the observed trend of decreasing $\epsilon$ towards the centres of
    clusters is evidence for physical effects in clusters causing fast rotator ETGs to have
    a lower average intrinsic ellipticity near the centres of rich clusters.

    \end{abstract}


\section{Introduction}\label{sec:introduction}

Early Type Galaxies (ETGs) account for half of the stellar mass in the local
Universe \citep{renzini2006}. They are traditionally divided in two subclasses:
Elliptical (E) and lenticular (S0) galaxies. To first order, Es have smooth,
single component
light profiles, while S0s present both a central bulge and an extended stellar
disc. ETGs are more common in clusters of galaxies, and the morphology-density
relation \citep[\TS, ][]{dressler1980} illustrates the effect of the local
environment on the formation and evolution of these galaxies
\citep[][argue however that the morphology correlates better with the
cluster-centric radius]{whitmore1993}.
\citet{dressler1997} used HST photometry to show that cluster Es are already in
place at redshifts $z \approx 0.5$, while the fraction of S0s is lower than in
the local Universe. Therefore Es form earlier than S0s, which
arise from infalling Late Type Galaxies (LTGs) mostly between $z \approx 0.5$ and $z =
0$ \citep[e.g. ][but see \citet{holden2009} for a different view]{vulcani2011}.\\

The division between Es and S0s presents however a number of problems.
Observationally it is difficult to distinguish Es from close-to-face-on S0s, and
morphological catalogues might be biased in this sense \citep{vandenbergh1990}.
Galaxies classified as Es are often found to contain disc components
\citep{kormendydjorgovski1989}. \citet{rixwhite1990} further demonstrated how discs can go
undetected even in local galaxies. All but the brightest Es and S0s in the
Coma Cluster form a family with continuous bulge-to-disc
light ratios \citep[B/D, ][]{jorgensen1994}. More recently, the \texttt{SAURON} survey
\citep{dezeeuw2002} used Integral Field Spectroscopy (IFS) to investigate the
stellar kinematics of ETGs. They identified two dynamical classes within ETGs:
Fast Rotators (FRs) and Slow Rotators (SRs). The former exhibit large scale
rotation patterns typical of a disc origin, while the latter have little to no rotation
\citep{emsellem2007, cappellari2007}. Importantly both SRs and FRs are found
amongst both Es and S0s. The volume-limited \atl survey \citep{cappellari2011a}
determined that 66\% of the local Es are FRs \citep{emsellem2011}, and a new
classification paradigm has been invoked \citep{cappellari2011b, kormendybender2012}.\\
In particular, FR ETGs form a sequence of increasing disc fraction, parallel to
that of spiral galaxies \citep[the comb diagram, ][]{cappellari2011b}. The emerging
picture is that - kinematically - FR ETGs are much more similar to spiral
galaxies than they are to SR ETGs. Unfortunately IFS observations, that are
necessary to tell apart FRs from SRs, are time consuming when compared to photometry.
Even the largest upcoming IFS surveys \citep[SAMI, MaNGA; ][]{croom2012, bundy2015}
have sample sizes which are several orders of magnitude smaller
than current state-of-the-art photometric surveys, like SDSS. Here we investigate
whether we can take advantage of currently available photometric samples to
investigate the properties of ETGs in the framework of the FRs/SRs classification.\\

Using photometry alone, ETGs can be characterised by their projected ellipticity
$\epsilon \equiv 1 -
q$, where $q$ is the apparent axis ratio. The observed value of \eps depends
both on the distribution of the orbits (which determines the \textit{intrinsic}
ellipticity) and on the inclination of the galaxy on the plane of the sky.
In general it is not possible to infer the 3-D structure of an individual galaxy
from its photometry alone, but we can study intrinsic shapes statistically
\citep{sandage1970, lambas1992}. These studies showed that the distribution of
ellipticity \distr for Es and S0s is different, with Es on average rounder than
S0s. More recently, \citet{weijmans2014} used the \atl sample of ETGs to show
that the intrinsic shape of FRs (both Es and S0s) is consistent with that of
spiral galaxies, with an average axis ratio $q = 0.25 \pm 0.01$. This reinforces
the view that FR ETGs and spiral galaxies form a family of intrinsically flat
stellar systems.

\citet[][KR05]{kuehnryden2005} studied the relation between $q$ and the local
environment using a magnitude limited sample ($r < 17.77 \; \mathrm{mag}$).
They found that galaxies with different magnitude and light profile
exhibit different trends of \eps with the local number density of galaxies.
Galaxies characterised by a de Vaucouleurs light profile \citep{devaucouleurs1948}
are rounder in denser environments, regardless of their luminosity. Galaxies
with exponential profiles show two opposite tendencies, based on their
absolute magnitude. Galaxies fainter than $M_r = -20 \; \mathrm{mag}$ are rounder
in denser environments, while more luminous galaxies tend to become flatter at
higher density.
These results suggest that the environment can affect the shape of galaxies.
Intriguingly, the trend observed by KR05 is strongest when the density is
measured inside an aperture with diameter $2 \, h^{-1} \mathrm{Mpc}$, which
roughly corresponds to the size of a cluster of galaxies.\\

Here we propose to investigate \distr for cluster galaxies, in the framework of
the SR/FR paradigm. In particular, since there are no SRs with $\epsilon \geq
0.4$ \citep{emsellem2011}, we can remove them from any photometric sample.
In the next section we introduce our sample of local cluster galaxies, and
proceed to show the results of the analysis. We then discuss possible sources of
bias, and whether the observed relation is a consequence of previously known
trends. We conclude with a summary of our results.

\section{Data and Sample}\label{sec:data.and.sample}

To assess the effect of the environment on the apparent ellipticity of galaxies
we study a sample of rich nearby clusters of galaxies with SDSS DR10 data
\citep{ahn2014}. At low redshift ($z \lesssim 0.04$), the most complete
catalogue of clusters is that of \citet{abell1989}, from which:

\begin{itemize}
  \item We discarded all the clusters with $z > 0.1$, or with no redshift
  measurement available.
  \item We discarded all clusters with richness $\mathcal{R} < 2$ 
  because we expect any signature of the cluster environment to be
  strongest in rich systems.
\end{itemize}

This left 20 clusters (see \reftab{tab:sample}). For each of them, we
retrieved SDSS photometry as follows.

\begin{itemize}
  \item We determined the centre of the cluster as the coordinates of the
  brightest galaxy. When the second brightest galaxy falls within $0.5 \;
  \mathrm{mag}$ from the brightest, we used the midpoint between the brightest
  and second brightest as the centre of the cluster (e.g. \Ab 1656, \Ab 1367).
  \item We queried the table ``Galaxy'' from the SDSS DR10 database, retrieving all
  the galaxies within a projected radius of 1.5 Mpc from the
  cluster centre. The angular distance was derived by assuming Planck
  Cosmology \citep[$\Omega_m = 0.32, \Omega_\Lambda = 0.68, h_0 =
  0.67$,][]{ade2014}. We assume zero peculiar velocity for all the clusters.
  \item We removed all galaxies with a radius \textit{deVRad}\footnote{
  Column names from the SDSS table are reported in italics.} $\leq 0.4''$
  (corresponding to 0.5 kpc at $z = 0.1$). This constraint filters out artefacts
  and measurements with bad photometry, which are otherwise still present in the
  table Galaxy.
  \item We further removed objects that are classified as stars in the $g'$-
  and $i'$-band (SDSS uses only $r'$-band for this classification). This condition
  filters out 31 objects. Objects with negative or zero errors in the photometry
  were also eliminated (21 objects in total).
  \item We applied a cut in absolute magnitude at $M_r = -18.0 \; \mathrm{mag}$. This is
  well above the $r'$-band completeness limit of SDSS at redshift $z = 0.1$
  (the maximum distance modulus for our sample of clusters is $\approx 38$ mag).
  A more generous magnitude cut increases the scatter in the RS, which
  is undesirable. We do not apply a k-correction, but this does not affect
  our results.
\end{itemize}

For each galaxy we retrieved the SDSS $r'$-band \textit{ModelMag} as a measure of
magnitude and $g'-r'$ colours using SDSS \textit{AperMag}.\\

Ancillary data consist of SDSS DR10 spectroscopic redshifts for a sample of
galaxies in the cluster \Ab 1656 (the Coma Cluster), and of Galaxy Zoo 2
\citep[GZ2, ][]{willett2013} morphological classifications, when available.

\subsection{Projected ellipticity}\label{subsec:projected.ellipticity}

SDSS offers two different measures of ellipticity. The model ellipticity is
defined as $1 - b/a$, where $b/a$ is the apparent axis ratio of the $r'$-band
best fit model (i.e. $b/a \equiv \textit{deVAB\_r}$ if \textit{fracDeV\_r}$
\geq 0.5$ and $b/a \equiv \textit{expAB\_r}$ if \textit{fracDeV\_r}$ < 0.5$).
The SDSS pipeline automatically corrects the model axis ratio for the effect of
the Point Spread Function (PSF; see \refsec{subsec:observation.bias} for a
discussion).

An alternative measure of the ellipticity is to use the adaptive second order
moments of the surface brightness $I$. This method uses a Gaussian
weight function adaptively matched to the size and shape of the galaxy being measured
\citep{bernsteinjarvis2002, hirataseljak2003}.

\begin{figure}
   \centering
   \includegraphics[type=pdf,ext=.pdf,read=.pdf,width=0.5\textwidth]{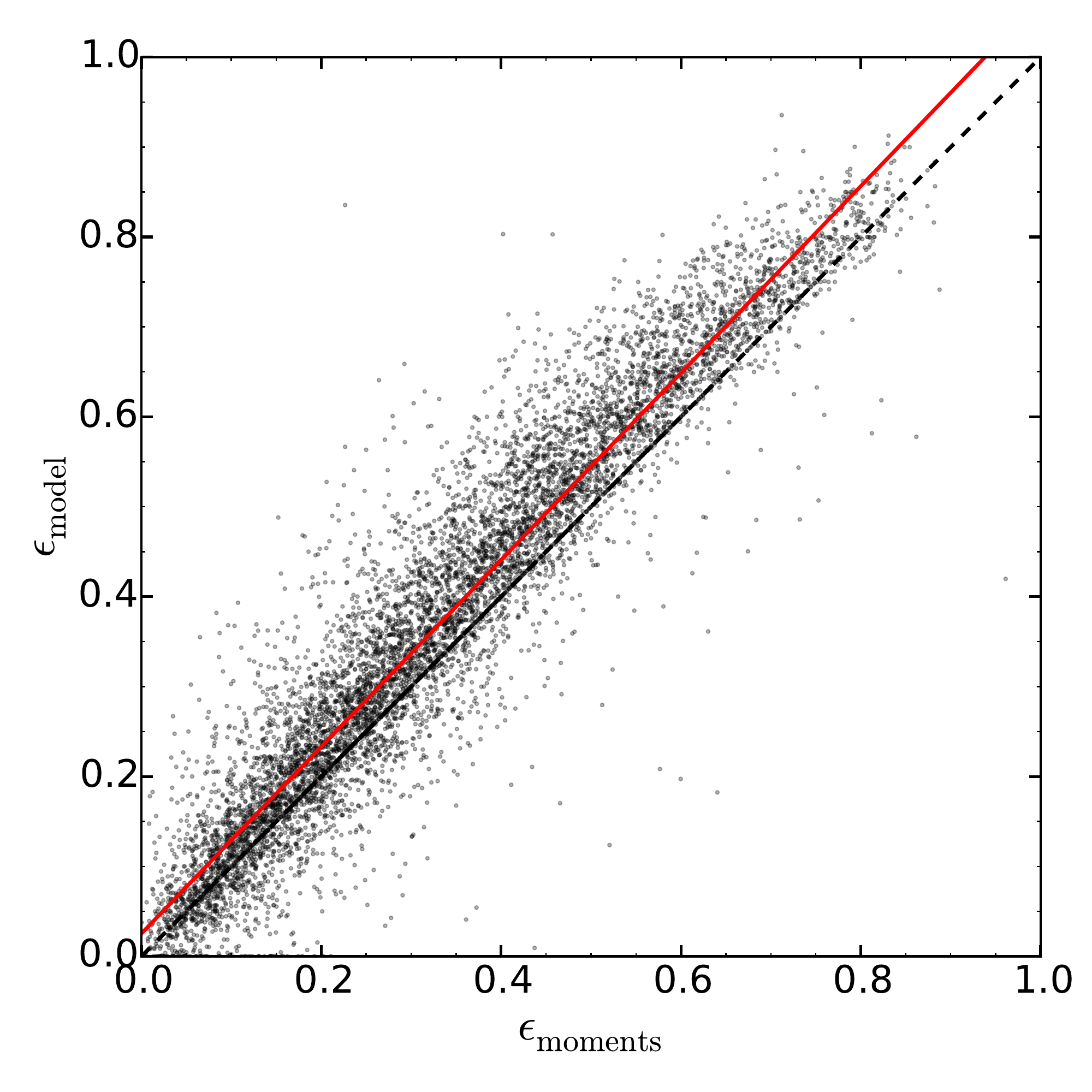}
   \caption{Comparison between $\epsilon_\mathrm{model}$ and
   $\epsilon_\mathrm{moments}$, measured in the $r'$-band as described in the
   text. The solid red line is the best-fit linear relation, while the black
   dashed line is the 1:1 relation. $\epsilon_\mathrm{model}$ is on average
   higher than $\epsilon_\mathrm{moments}$.}\label{fig:eps.mod.vs.mom}
\end{figure}

Comparing the model ellipticity to the moments ellipticity in the $r'$-band, we find that in
general the former is larger. The best linear fit yields $\epsilon_\mathrm{model} =
(1.039 \pm 0.004) \epsilon_\mathrm{moments} + (0.025 \pm 0.001)$ (\reffig{fig:eps.mod.vs.mom}).
As in general \eps increases with radius within galaxies, $\epsilon_\mathrm{model}$
is better suited for the outer region of galaxies.
This is reinforced by the fact that most of the cases where
$\epsilon_\mathrm{moments} \gg \epsilon_\mathrm{model}$ are close-to-face-on
barred galaxies.
We are primarily interested in the effect of the cluster environment on the
shape of galaxies. Since this effect is larger at larger galactic radius, we
choose to use $\epsilon_\mathrm{model}$ as a measure of the shape of galaxies.
See Appedix \ref{app:moments.ellipticity} for a comparison with the results using
$\epsilon_\mathrm{moments}$.\\

We investigate the variation of \eps as a function of
two different tracers of the environment density, the projected cluster-centric
radius $R$ and the projected number density of galaxies \sig3. The latter was
measured for each galaxy inside the circle on the sky comprising its three
closest neighbours \citep[e.g. ][]{cappellari2011b}.

\subsection{Red Sequence determination}

For each cluster we constructed the $g'-r'$ vs $r'$ Colour-Magnitude Diagram
(CMD). In order to reject any outliers and artefacts we eliminated all the points
in the CMD with $g'-r' \notin [0.0, 2.5]$ and galaxies with
errors in $g'-r'$ greater than $0.1$.

We identified the Red Sequence (RS) using the Gaussian mixture
model of \citet{houghton2012}. The algorithm fits two superimposed
Gaussian models: one for the RS and one for the underlying galaxy distribution,
including background and foreground objects.
The mean $\langle g'-r' \rangle$ of the RS Gaussian is allowed to vary linearly with the
magnitude $r'$, i.e. $\langle g'-r' \rangle = m \, (r' - 16) + c$, where $m$ and $c$ are
constants to be determined. The amplitude and the dispersion $\Delta$ of the
Gaussian are held constant with $r'$.
For the background Gaussian, all parameters are independent of $r'$.

The algorithm determines the most likely values of $m$, $c$ and $\Delta$. In addition, it
returns the probability $P(\rm RS)$ that each galaxy on the CMD belongs to the
RS, and we define galaxies with $P(\rm RS) > 0.05$ to be members of the RS. This
corresponds to a $2 \sigma$ selection.
After applying this procedure to all the 20 clusters in the sample, the average
results are: $\langle m \rangle = -0.037 \pm 0.006$ and $\langle \Delta \rangle
= 0.035 \pm 0.007$.
 
\begin{figure}
   \includegraphics[width=0.53\textwidth]{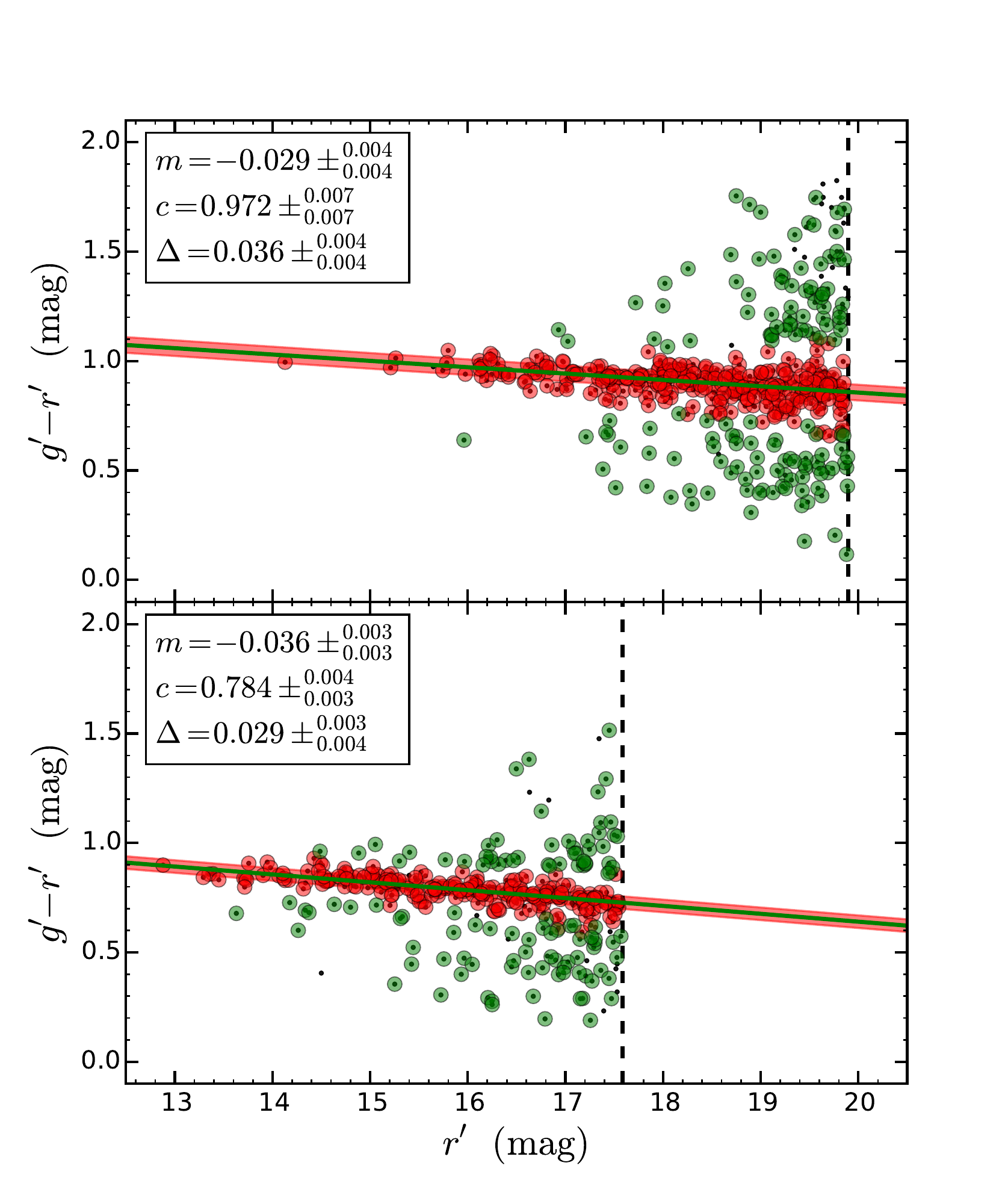}
   \caption{Two example CMDs for \Ab 1650 at $z = 0.08$ (top) and \Ab 2199 at
   $z = 0.03$ (bottom). Galaxies are marked
   by black dots. Red and Green circles are overlaid on RS galaxies and to non-RS
   galaxies. Naked dots denote galaxies with large errors that were not used in
   the fit. The best fit to the RS is the green line, while the region shaded in
   red represents the intrinsic scatter. The vertical dashed line is the
   adopted magnitude cut. The best fit parameters are reported in the top left
   box ($m$ is the slope, $c$ the intercept and $\Delta$ the intrinsic scatter).}
   \label{fig:cmd.pair}
\end{figure}

In \reffig{fig:cmd.pair} we show the CMD of two clusters, \Ab 1650 at $z = 0.08$
(top panel) and \Ab 2199 at $z = 0.03$ (bottom panel). Each black dot represents
a galaxy from the SDSS catalogue,
but only the dots with an overlaid circle have been considered for the purpose
of the RS determination. Red circles mark RS galaxies, while green circles mark
non-RS galaxies, as determined by their value of $P(\rm RS)$. The vertical
dashed line is the magnitude cut for the cluster, while the solid green line
represents the best fit to the RS, i.e. the line of equation $g'-r' = m \, (r' -
16) + c$. In the top left box of each panel we report the best fit value of $m$,
$c$ and $\Delta$ with the corresponding 1 $\sigma$ confidence interval.\\

We constructed a Reference Cluster (RC) as the union of all the clusters in the
sample, and further constructed a reference RS as the union of all the RS's of
all the clusters. The RC consists of 9052 galaxies, the reference
RS sample counts 5175 entries (see last row in \reftab{tab:sample}, columns 7 and 8).

\subsection{Redshift selection}\label{subsec:redshift.selection}

A colour selection can be useful to reject interlopers, but it is not as
reliable as a spectroscopic selection. Individual cluster galaxies can be
significantly bluer than allowed by the RS (for instance, if they are undergoing
star formation). On the other hand, a number of physical processes can affect
the colour of interlopers, placing them close or on the RS (dust reddening,
star formation). In order to assess the number of interlopers still present in
the RS sample, we used spectroscopic data from SDSS DR10. Unfortunately only \Ab
1656 had enough candidate galaxies with spectra, therefore
the spectroscopic sample is limited to one cluster. We retrieved the redshift
$z$ for all the galaxies in the photometric sample, and rejected all the
galaxies with $z < 0.01$ or $z > 0.04$ \citep{price2011}. The sample for \Ab
1656 is reduced from 468 to 387 galaxies, while the RS is reduced from
362 to 333 members. The number of interlopers is 61 and 21 for the full
sample and for the RS respectively, corresponding to a fraction of 0.14 and 0.06.
We then assumed that all of the galaxies with no redshift measurement in the
database are interlopers too (20 and 8 galaxies for the full sample and for the
RS respectively). This gives a maximum fraction of interlopers of 0.18 and 0.08.

\section{Results}\label{sec:results}

We start this section studying if \distr varies between different clusters, and
the properties of the RS sample (\refsec{subsec:full.sample}). We then
study the redshift selected sample of \Ab 1656 to study the effect of
interlopers (\refsec{subsec:spectroscopic.a1656}). In
\refsec{subsec:flat.galaxies} we consider again the full sample, where we repeat
the analysis for intrinsically flat galaxies. We then study the relation between
\eps and $r'$-band luminosity (\refsec{subsec:dependence.on.luminosity}) and
conclude this section by studying a subsample of morphologically selected
galaxies (\refsec{subsec:galaxy.zoo2}) and splitting our sample according to the
shape of the luminosity profile (\refsec{subsec:luminosity.profile}).

\subsection{Full sample}\label{subsec:full.sample}

\begin{figure}
   \centering
   \includegraphics[type=pdf,ext=.pdf,read=.pdf,width=0.5\textwidth]{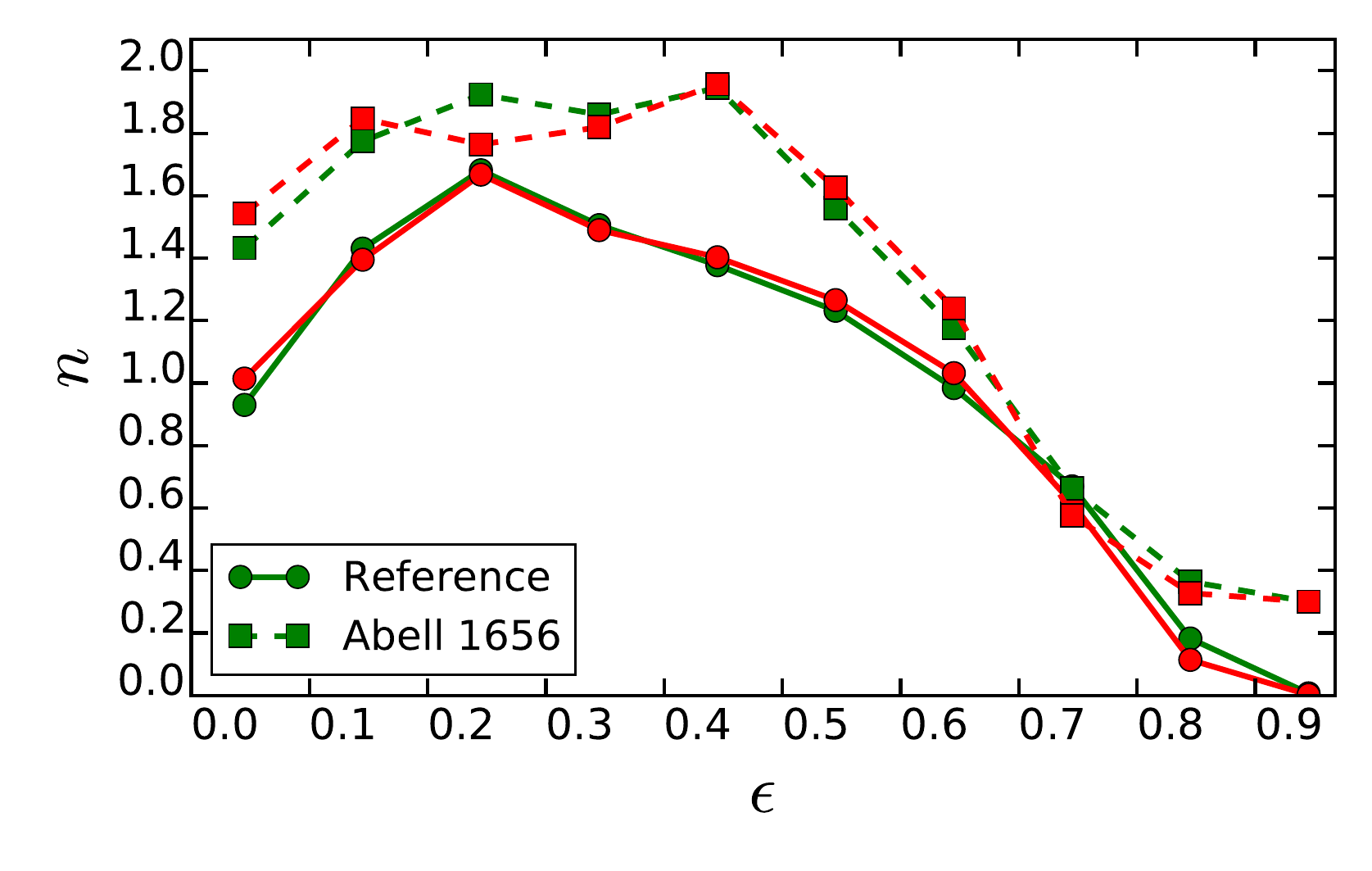}
   \caption{Distribution of projected \eps for the Reference Cluster (solid
   green line) and the reference RS (solid red line). The dashed lines show
   the distribution for \Ab 1656, with the same colour coding. According to a KS
   test (\reftab{tab:sample}), \Ab 1656 has a different \eps distribution than
   the Reference Cluster. This is true also for the RS. The distributions are
   normalised so that the integral is unity, but the values of $n$ for \Ab 1656
   are plotted 0.3 higher for display purposes.}\label{fig:eps.distr}
\end{figure}

In \reffig{fig:eps.distr} we plot \distr of the RC (solid
green line), alongside the distribution of the reference RS (solid red line).
Each distribution has been normalised so that its integral between $\epsilon =
0$ and $\epsilon = 1$ is unity. For both the RC and the RS we
find an average ellipticity $\langle \epsilon \rangle = 0.38 \pm 0.01$ and a
standard deviation $\sigma_\epsilon = 0.21 \pm 0.01$.

We then used a Kolmogorov-Smirnoff (KS) test on the null hypothesis that
any individual cluster in the sample has the same \distr as the RC,
and that any of the cluster RS has the same \distr as the reference RS. The
results are summarised in \reftab{tab:sample}. For each cluster we list the
number of galaxies found inside the adopted SDSS aperture (column 7) and the
number of galaxies on the RS (column 8). We then report the
results of the KS test: for each cluster we list the probability that its \distr
is the same as the RC (column 11), and that \distr of its RS is the same
as the reference RS (column 12). We have only one entry with $P \lesssim 0.05$
for both column 11 and 12 (\Ab 1656 and \Ab 16 respectively).
If we assume that \distr is the same between all clusters,
we expect the results of the KS test to be distributed uniformly between 0
and 1. We performed an Anderson-Darling \citep[AD, ][]{andersondarling1954}
uniformity test for the null hypothesis that the values of columns 11 and 12
of \reftab{tab:sample} are drawn from the uniform distribution. The AD values
are 0.48 and 0.41 for columns 11 and 12 respectively, corresponding to p-values
$\gg 0.25$ \citep{rahman2006}. Therefore there is no evidence of any difference
between the \distr of our sample of clusters.\\

Next we studied the variation of \eps with the cluster-centric radius $R$.

\begin{figure}
   \centering
   \includegraphics[type=pdf,ext=.pdf,read=.pdf,width=0.5\textwidth]{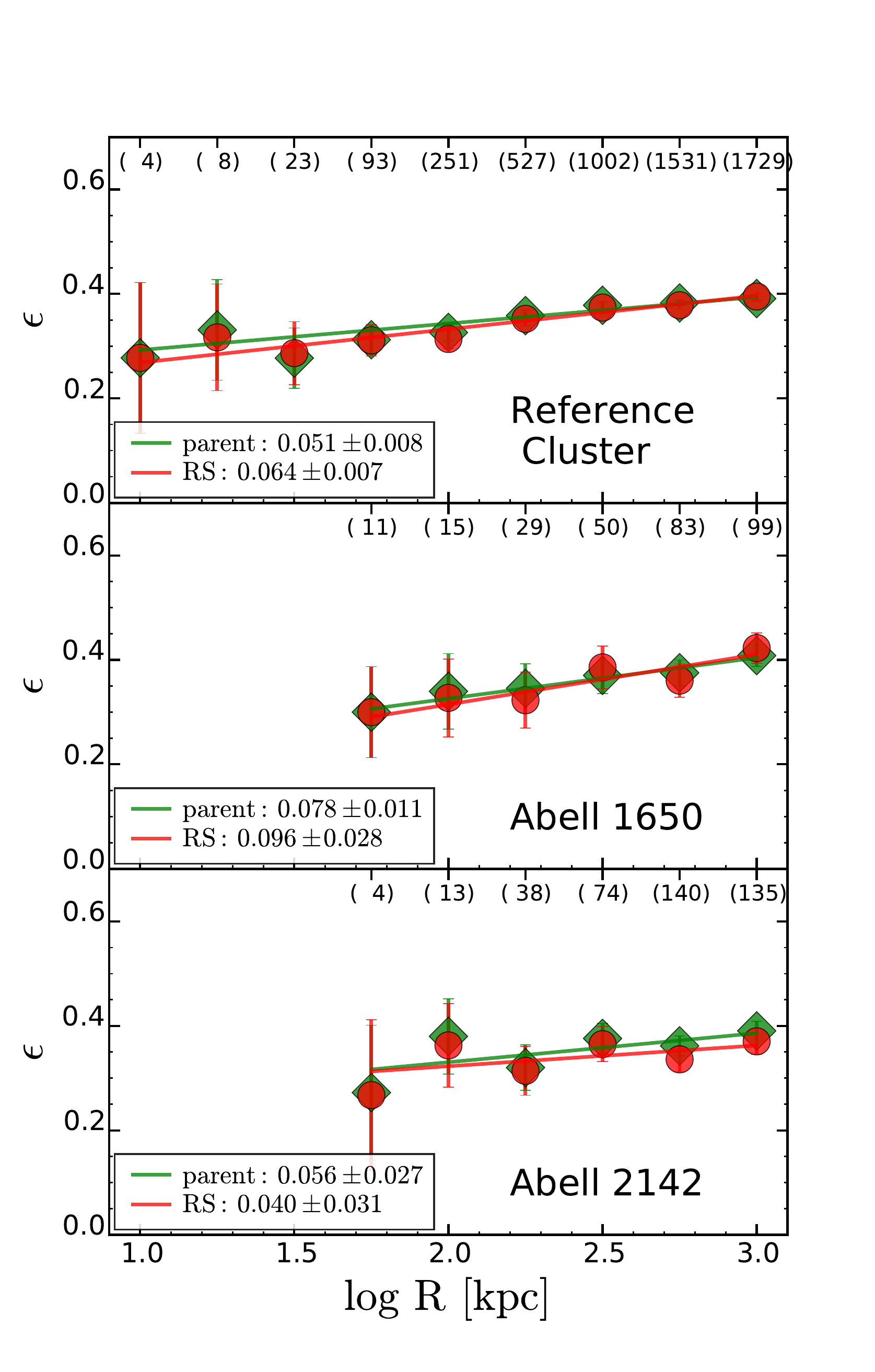}
   \caption{The projected ellipticity \eps of galaxies as a function of the
   projected cluster-centric radius. The Reference Cluster is
   shown at the top, followed by two example clusters: \Ab 1650 (centre) and \Ab
   2142 (bottom). The green diamonds represent all galaxies in the cluster; red
   circles represent RS galaxies only. The boxes in the bottom left corners
   report the best fit slope to the points and associated errors. The reference
   sample, as well as \Ab 1650, show a trend of \eps with \logR, which we call
   the \epsR relation. In \Ab 2142 the trend is not significant (the slope is less
   than three standard deviations above 0). The numbers in parentheses are the number of
   galaxies in each bin of \logR for the RS.}
   \label{fig:eps.vs.logr}
\end{figure}

In \reffig{fig:eps.vs.logr} we plot \eps for nine bins in \logR. The green
diamonds mark each bin from the parent sample in the CMD, and the red
circles mark \eps for the galaxies on the RS. At the top of each panel, the
numbers in parentheses are the numbers of RS galaxies in each bin (we eliminated
any bin with less than three galaxies). The errorbars
were derived assuming that for each bin in \logR, the distribution \distr is
flat between $\epsilon = 0$ and $\epsilon = 1$, and zero otherwise. Such a
distribution has standard deviation $1 / \sqrt{12}$, therefore the error on the
mean \eps for each bin is $1 /\sqrt{12 \; N}$, where $N$ is the number of
galaxies in that bin. Since \distr is hardly flat, and does not extend to
$\epsilon = 1$ (\reffig{fig:eps.distr}), it follows that our estimate of the
error is a conservative one.
The box in the bottom left corner reports the slope (with errors) of the best
linear fit, using least squares minimization. The uncertainties on the slopes
were derived from the errorbars, therefore they too represent a conservative estimate.
The top panel depicts the results for the RC: both the parent and
RS sample show a clear trend of \eps with \logR, in that the best-fit slope is
more than three standard deviations away from 0. For brevity, we call the
observed trend the ellipticity-radius relation (\eps-$R$). The central panel
shows the same plot for \Ab 1650, a cluster where the \epsR trend is clearly
present. In contrast, \Ab 2142 (bottom panel) is an example of a cluster where
the relation is not detected, i.e. the measured slope is within three standard
deviations of 0. For the parent samples, four clusters have a slope that is more
than three $\sigma$ from 0, including \Ab 1650. Nine clusters have slopes between
one and three $\sigma$ above 0, and seven clusters (including \Ab 2142) have a
slope that is within one $\sigma$ from 0 (this includes \Ab 16 and \Ab 168, which
have negative slopes). The breakdown is 4/9/7. For the RS samples the breakdown
is 5/10/5 (\Ab 16 and \Ab 168 still have negative slopes). In all clusters the
best-fit slope of the parent sample and of the RS sample are statistically
consistent to the level of three $\sigma$.

To estimate the probability that our result arises from chance, we assume
Gaussian errors on the best-fit slope and use the binomial distribution:
\begin{equation}\label{eq:binom}
    f(k; n, p) \equiv \binom{n}{k} p^k (1 - p)^{n-k}
\end{equation}
The probability that out of 20 best-fit slopes 13 are more than one sigma
above 0 is $f(13, 20, 0.16) \approx 1 \times 10^{-6}$. We therefore
conclude that the observed \epsR relation is a real effect.\\

We find no correlation between the value of the best fit slope and global
cluster parameters, like X-ray temperature or X-ray luminosity
\citep{ebeling1996}, Richness \citep{abell1989} or Bautz-Morgan type
\citep{bautzmorgan1970}.

\subsection{Spectroscopic sample for \Ab 1656}\label{subsec:spectroscopic.a1656}

\begin{figure}
   \centering
   \includegraphics[type=pdf,ext=.pdf,read=.pdf,width=0.5\textwidth]{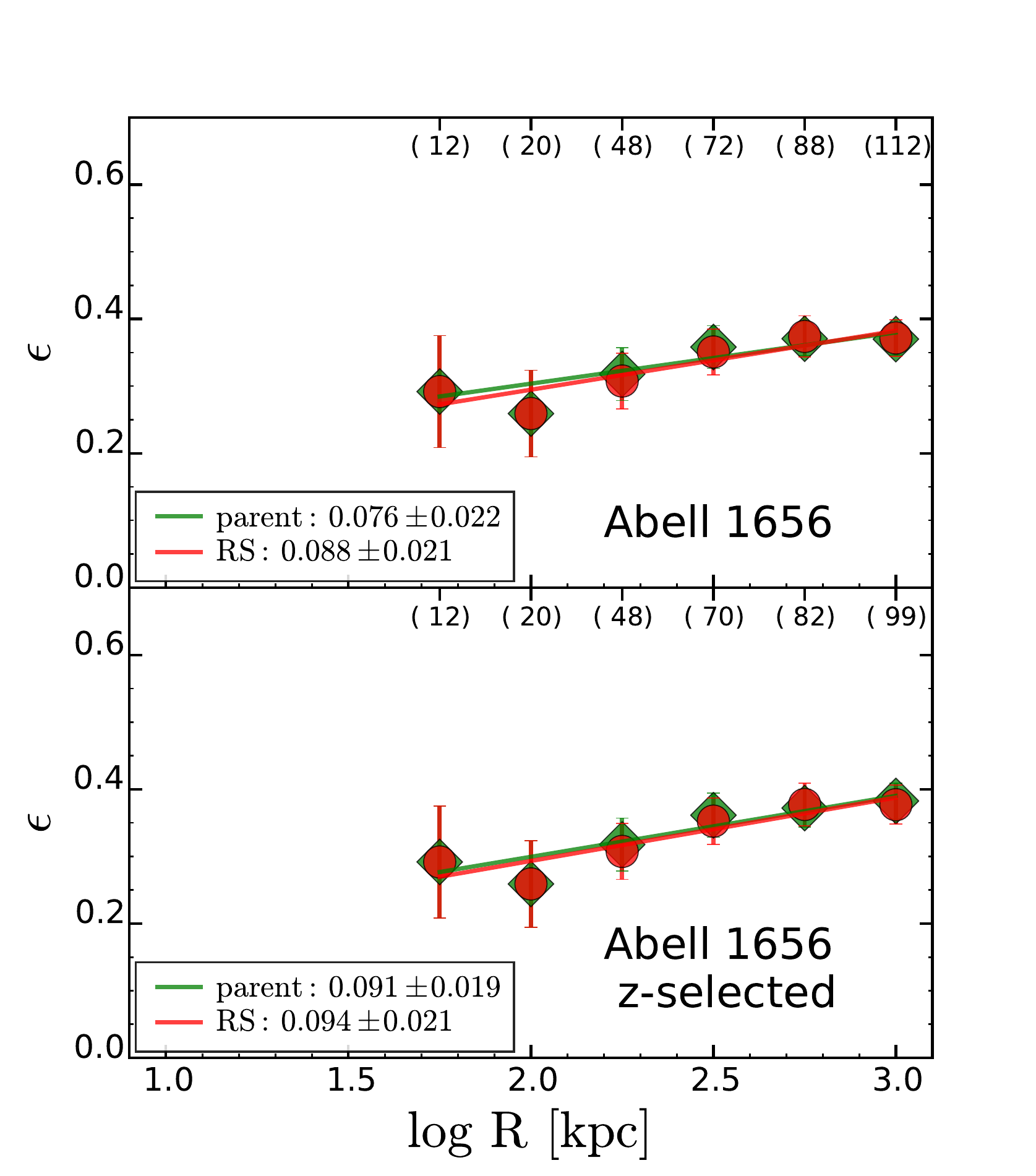}
   \caption{Same as \reffig{fig:eps.vs.logr}, for the \Ab 1656 sample (top
   panel) and for the sample of redshift selected cluster members
   (bottom panel; see \refsec{subsec:redshift.selection}). The trend of \eps
   with cluster centric radius $R$ observed in \Ab 1656 is not an effect of
   interlopers: the slope of the best-fit linear relations are statistically
   consistent between the original sample and the redshift selected samples
   (boxes in the bottom left corner of each panel).}\label{fig:eps.vs.logr.coma}
\end{figure}

In \refsec{subsec:redshift.selection} we presented a sample of redshift selected
cluster member for \Ab 1656. The \epsR relation stays the same for the original
sample (\reffig{fig:eps.vs.logr.coma}, top panel) and for the redshift selected
sample (\reffig{fig:eps.vs.logr.coma}, bottom panel). The significance of the
trend increases for both the parent sample and the RS, but the trends are
statistically consistent. If anything, the slope of the best-fit relation
increases going from the original sample to the spectroscopic sample.

\subsection{Flat galaxies}\label{subsec:flat.galaxies}

We repeated the analysis for the subset of flat galaxies. \citet{emsellem2011}
showed that there are no SRs flatter than $\epsilon \gtrsim 0.4$
\citep[we treat double sigma galaxies as FRs,][]{krajnovic2011}. Therefore we
isolate a sample of intrinsically flat galaxies by selecting $\epsilon \geq
0.4$\footnote{Notice that this biases our sample against FRs with lower intrinsic
ellipticity, because the range of possible inclinations that satisifies the constraint
$\epsilon \geq 0.4$ decreases with decreasing intrinsic ellipticity.}.
We performed a KS test comparing the flat galaxies in the RC to those in
each single cluster. We then repeated the test for
RS galaxies only. The results are listed in \reftab{tab:sample} (columns 13
and 14). We find that \Ab 1656 fails both tests and \Ab 1650 fails one test.
To test the null hypothesis that the values in columns 13 and 14 are distributed
uniformly between 0 and 1, we use again the AD uniformity test. The resulting
values are 0.53 and 1.84 for columns 13 and 14 respectively, corresponding to
p-values $\gg 0.25$ and $0.10 < \mathrm{p-value} < 0.15$. We infer that there
is no conclusive evidence to suggest that the distribution of \eps is different
between different clusters, as we cannot reject the null hypothesis with
confidence greater than $P = 0.01$.

However, by looking at \Ab 1656 in \reftab{tab:sample}, we see that this cluster
scores very low confidence in three out of four tests (columns 11, 13 and 14).
Since these values are not independent, as a conservative estimate we adopt the
largest one as the probability that the $n(\epsilon)$ of Abell 1656 is the same
as the $n(\epsilon)$ of the RC, $P(\mathrm{same}) \leq 0.05$.\\

\begin{figure}
   \centering
   \includegraphics[type=pdf,ext=.pdf,read=.pdf,width=0.5\textwidth]{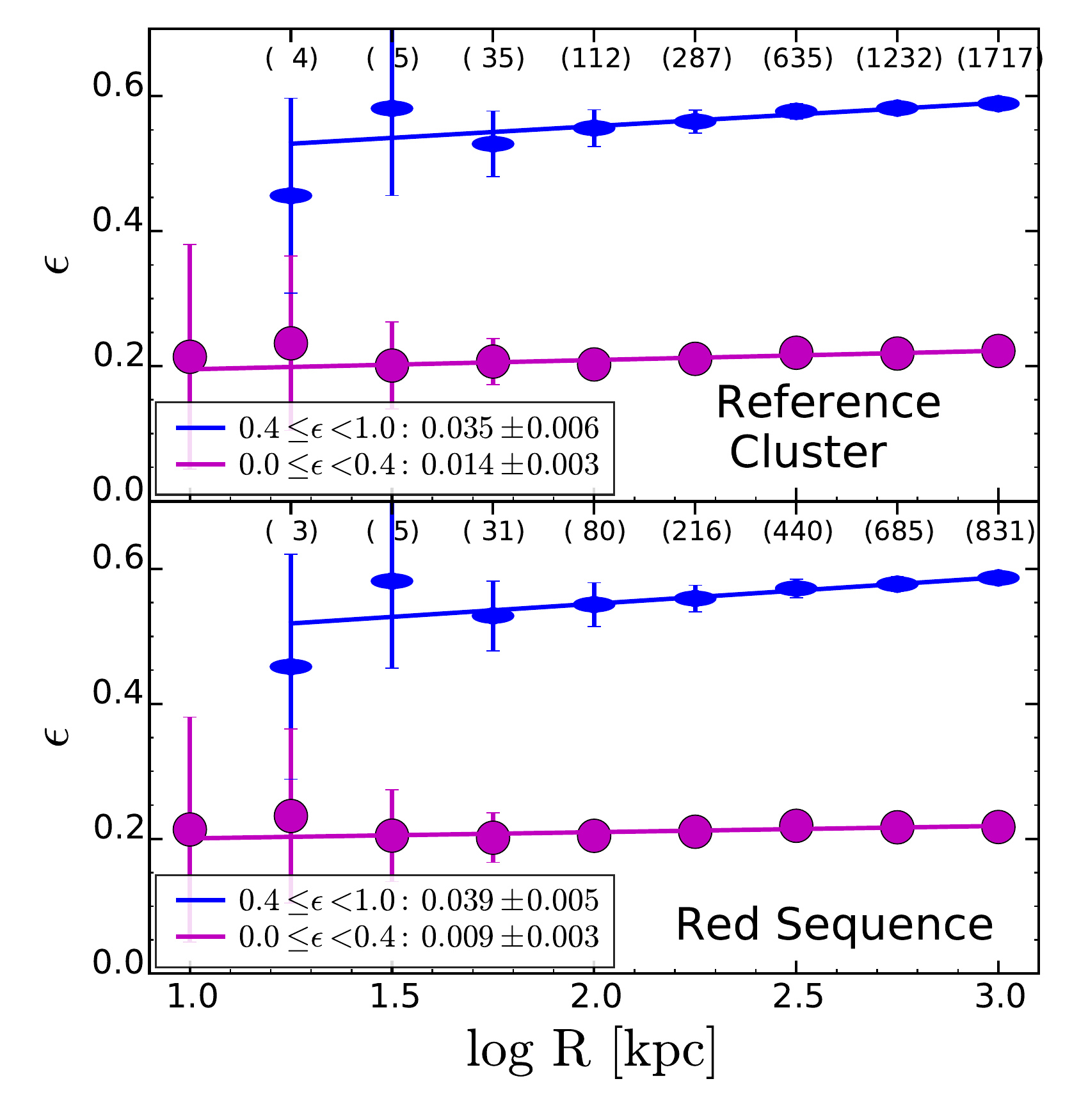}
   \caption{The projected ellipticity \eps of galaxies as a function of the
   projected cluster-centric radius. The Reference Cluster is shown at the top,
   the reference RS sample in the bottom panel. In each plot, the blue ellipses represent
   galaxies that appear flat on the sky ($\epsilon \geq 0.4$). The magenta circles
   represent galaxies that appear round on the sky ($\epsilon < 0.4$).
   The blue and magenta lines are least squares best fits to the data; their
   slopes are reported in the bottom left corner of each figure. The number in
   parentheses are the number of galaxies in each bin of \logR for the flat
   subsample.}\label{fig:eps4.vs.logr}
\end{figure}

We now look into the relation of \eps with the cluster centric radius $R$.
The top panel of \reffig{fig:eps4.vs.logr} shows the \epsR relation for the
subsample of flat galaxies. The magenta circles are galaxies
with $\epsilon < 0.4$ (binned in \logR), and the blue ellipses are galaxies
with $\epsilon \geq 0.4$. The trend is observed for both the round and flat
subsamples, however the \epsR relation for flat galaxies ($0.035 \pm 0.005$) is
both steeper and and statistically more significant than that of round galaxies
($0.014 \pm 0.003$).
The bottom panel of \reffig{fig:eps4.vs.logr} shows the same relation for
galaxies on the RS: the best fit slope for flat galaxies is consistent with
the previous one. The change in mean \eps for flat galaxies is $\Delta \,
\epsilon \approx 0.1$ over the observed range in \logR.

If we repeat our analysis substituting \logR with \logsig3, we find
equivalent results: \eps decreases with \logsig3. This is a consequence of
the anticorrelation between R and \sig3. At this stage it is not possible to disentangle
which relation is more fundamental.

\begin{table*}
\centering
\setlength{\tabcolsep}{4pt}
\caption[]{The local sample of clusters and results of the KS tests.}\label{tab:sample}
\begin{tabular}{cccccccccccccc}
\hline
 ACO & RA & DEC & z & r & R &
 $\mathrm{N_{parent}}$  &  $\mathrm{N_{RS}}$  &
 $\mathrm{N_{\epsilon \geq 0.4}}$ & $\mathrm{N_{RS, \epsilon \geq 0.4}}$ &
 $\mathrm{P_{parent}}$ &
 $\mathrm{P_{RS}}$ &
 $\mathrm{P_{\epsilon \geq 0.4}}$ &
 $\mathrm{P_{RS, \epsilon \geq 0.4}}$ \\
\hline
  & (J2000) & (J2000) & & (mag) & (arcsec) & & & & & & & & \\
 (1)  &  (2)  &  (3)  &  (4) &
 (5)  &  (6)  &  (7)  &  (8) &
 (9)  &  (10) &  (11) &  (12)&
 (13) &  (14) \\
\hline

     16 & 0:16:46.30  & 6:44:39.84   & 0.084 & 19.88 & 16.09 & 416 & 185 & 187 & 99  & 0.875 & \textbf{0.031} & 0.752 & 0.656 \\
    168 & 1:15:9.79   & 0:14:50.64   & 0.045 & 18.48 & 28.52 & 330 & 181 & 148 & 84  & 0.959 & 0.846 & 0.854 & 0.555 \\
   1035 & 10:32:7.20  & 40:12:33.12  & 0.080 & 19.77 & 16.80 & 461 & 194 & 207 & 94  & 0.574 & 0.459 & 0.147 & 0.212 \\
   1186 & 11:13:51.36 & 75:23:39.84  & 0.079 & 19.75 & 16.95 & 311 & 152 & 128 & 63  & 0.395 & 0.950 & 0.533 & 0.408 \\
   1190 & 11:11:46.32 & 40:50:41.28  & 0.079 & 19.75 & 16.89 & 358 & 224 & 162 & 112 & 0.754 & 0.159 & 0.365 & 0.211 \\
   1367 & 11:44:29.52 & 19:50:20.40  & 0.021 & 16.83 & 58.29 & 284 & 206 & 118 & 87  & 0.161 & 0.219 & 0.149 & 0.076 \\
   1650 & 12:58:46.32 & -1:45:10.80  & 0.085 & 19.90 & 15.97 & 461 & 290 & 220 & 138 & 0.239 & 0.649 & \textbf{0.006} & 0.117 \\
   1656 & 12:59:48.72 & 27:58:50.52  & 0.023 & 16.99 & 54.13 & 468 & 362 & 197 & 153 & \textbf{0.046} & 0.105 & \textbf{0.004} & \textbf{0.010} \\
   1775 & 13:41:55.68 & 26:21:53.28  & 0.072 & 19.54 & 18.38 & 456 & 234 & 202 & 102 & 0.824 & 0.962 & 0.714 & 0.692 \\
   1795 & 13:49:0.48  & 26:35:6.72   & 0.062 & 19.20 & 21.14 & 421 & 220 & 176 & 85  & 0.240 & 0.340 & 0.307 & 0.633 \\
   1904 & 14:22:7.92  & 48:33:22.32  & 0.071 & 19.49 & 18.76 & 442 & 252 & 204 & 118 & 0.902 & 0.314 & 0.912 & 0.233 \\
   2029 & 15:10:58.80 & 5:45:42.12   & 0.077 & 19.68 & 17.43 & 558 & 285 & 260 & 129 & 0.381 & 0.825 & 0.535 & 0.578 \\
   2065 & 15:22:42.72 & 27:43:21.36  & 0.072 & 19.54 & 18.40 & 567 & 305 & 263 & 140 & 0.414 & 0.483 & 0.206 & 0.424 \\
   2142 & 15:58:16.08 & 27:13:28.56  & 0.090 & 20.04 & 15.10 & 646 & 405 & 281 & 158 & 0.708 & 0.092 & 0.442 & 0.426 \\
   2151 & 16:5:14.88  & 17:44:54.60  & 0.037 & 18.04 & 34.41 & 355 & 189 & 168 & 86  & 0.305 & 0.460 & 0.254 & 0.143 \\
   2199 & 16:28:36.96 & 39:31:27.48  & 0.030 & 17.59 & 41.80 & 327 & 201 & 145 & 80  & 0.847 & 0.527 & 0.827 & 0.529 \\
   2244 & 17:2:43.92  & 34:2:48.48   & 0.099 & 20.27 & 13.82 & 556 & 301 & 247 & 133 & 0.573 & 0.873 & 0.172 & 0.361 \\
   2255 & 17:12:30.96 & 64:5:33.36   & 0.081 & 19.80 & 16.61 & 636 & 398 & 294 & 186 & 0.346 & 0.262 & 0.727 & 0.797 \\
   2256 & 17:3:43.44  & 78:43:2.63   & 0.060 & 19.12 & 21.82 & 528 & 341 & 217 & 138 & 0.317 & 0.137 & 0.752 & 0.412 \\
   2670 & 23:54:10.08 & -10:24:18.00 & 0.076 & 19.66 & 17.56 & 471 & 250 & 206 & 109 & 0.474 & 0.676 & 0.974 & 0.169 \\

\hline

   Reference  & - & - & - & - & - & 9052  & 5175 & 4030 & 2294 & - & - & - & - \\

\hline
\end{tabular}

Column (1): cluster ID \citep{abell1989}.
Column (2): right ascension in degrees and decimal.
Column (3): declination in degrees and decimal.
Column (4): cluster redshift from \citet{ebeling1996}, or from \citet{abell1989} when available.
Column (5): magnitude cut adopted, SDSS $r$-band.
Column (6): cut in the projected distance, corresponding to 1.5 Mpc.
Column (7): number of galaxies in the parent sample.
Column (8): number of galaxies in the RS.
Column (9): number of flat ($\epsilon \geq 0.4$) galaxies in the parent sample.
Column (10): number of flat ($\epsilon \geq 0.4$) galaxies in the RS.
Column (11): probability that the parent sample has the same \eps distribution as
the corresponding Reference Cluster.
Column (12): probability that the RS has the same \eps distribution as
the corresponding Reference Cluster.
Column (13): probability that the sample of flat ($\epsilon \geq 0.4$) galaxies has
the same \eps distribution as the corresponding Reference Cluster.
Column (14): probability that the sample of flat ($\epsilon \geq 0.4$) RS galaxies
has the same \eps distribution as the corresponding Reference Cluster.
Probability values $P < 0.05$ are highlighted in boldface characters (see
\refsec{subsec:full.sample} and \refsec{subsec:flat.galaxies}).

\end{table*}

\subsection{Dependence on luminosity}\label{subsec:dependence.on.luminosity}

\begin{figure}
   \centering
   \includegraphics[type=pdf,ext=.pdf,read=.pdf,width=0.5\textwidth]{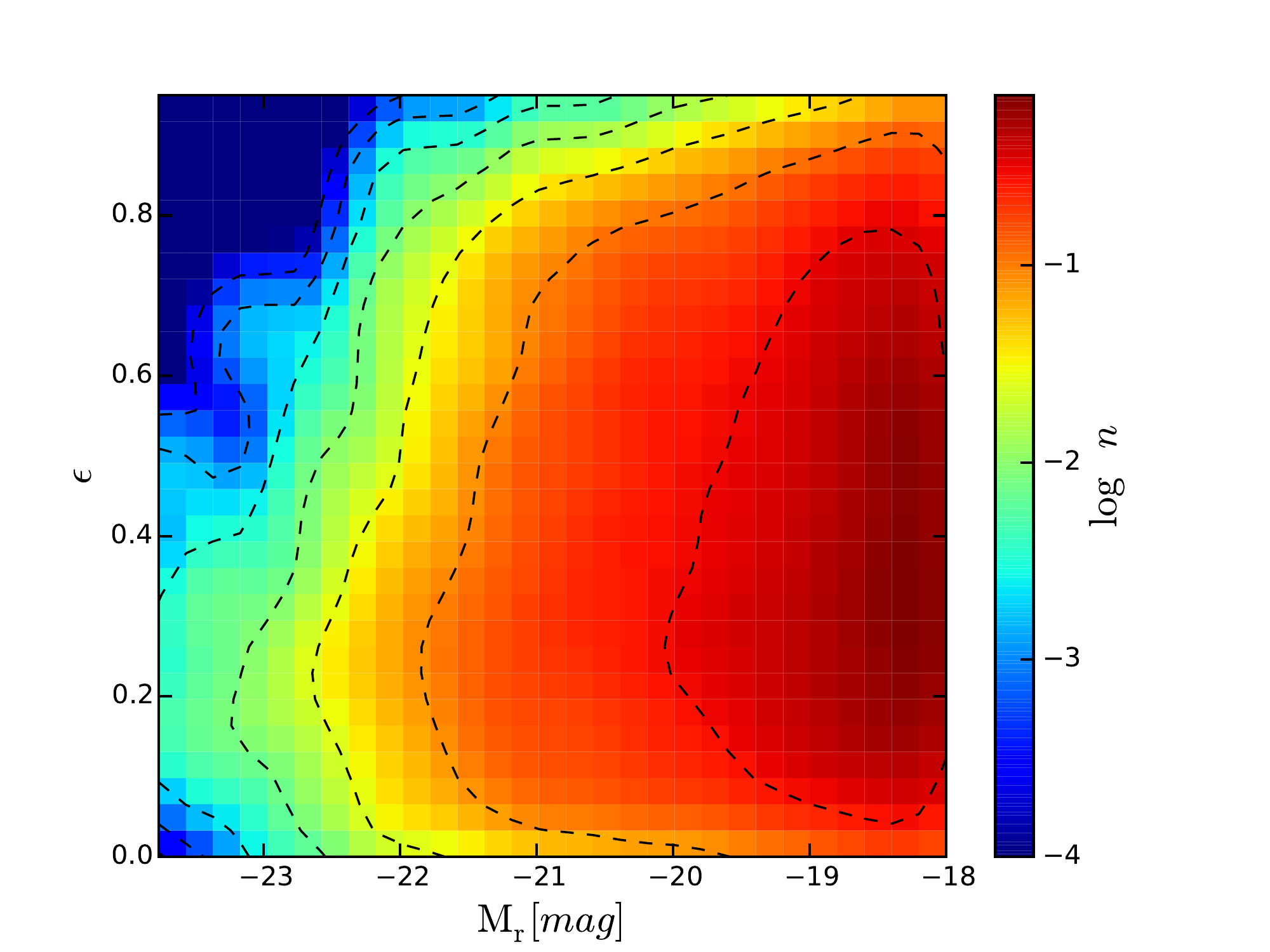}
   \caption{The distribution of the galaxies (Reference Cluster) in the
   magnitude-\eps plane. $n$ is the number density of galaxies in the $\epsilon - 
   \mathrm{M_r}$ space. There is an excess of round, luminous galaxies (\Mr $\lesssim
   -22 \; \mathrm{mag}$, bottom left corner).}\label{fig:eps.vs.Mr}
\end{figure}

In \reffig{fig:eps.vs.Mr}
we show how the RC populates the \eps vs \Mr-space. The colourbar
represents the $\log$ number density of galaxies. 
We notice that there is an excess of round, luminous galaxies at \Mr $\approx
-23 \; \mathrm{mag}$ (\reffig{fig:eps.vs.Mr}, bottom left corner). In contrast,
there are only 31/9052 galaxies with $\epsilon \geq 0.4$ and \Mr $\leq -22 \; \mathrm{mag}$, and
only 2/9052 galaxies with $\epsilon \geq 0.4$ and \Mr $\leq -23 \; \mathrm{mag}$. It is possible
that a number of these are foreground interlopers. The brightest of
these galaxies is however the cD galaxy of \Ab 1650, which is genuinely flat.
For \Mr $\lesssim -22 \; \mathrm{mag}$, the contour lines indicate a gradual
increase of \eps with decreasing $r'$-band luminosity.

\begin{figure}
    \centering
    \includegraphics[type=pdf,ext=.pdf,read=.pdf,width=0.5\textwidth]{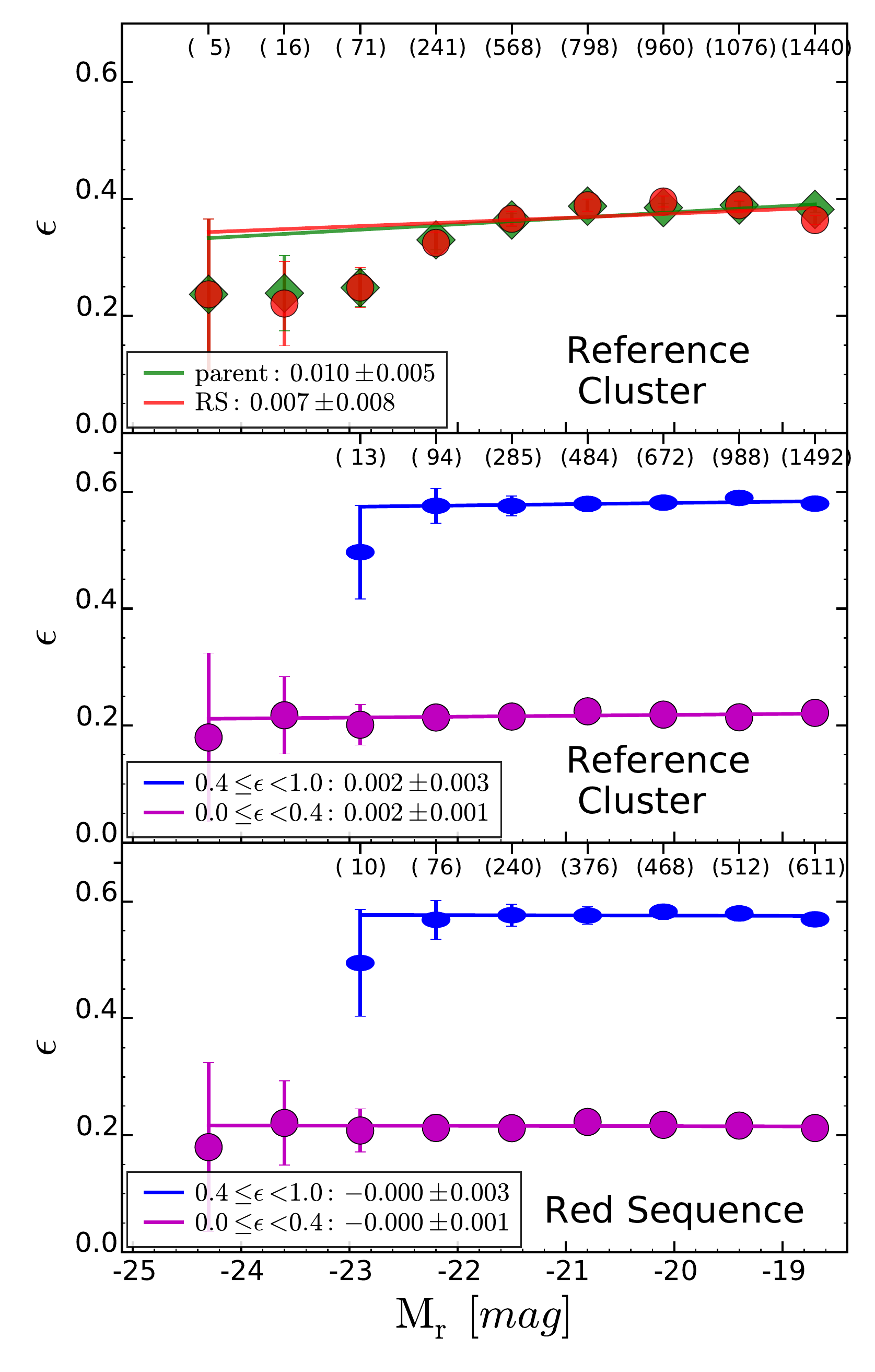}
    \caption{Top panel: \eps vs \Mr for the Reference Cluster (green diamonds)
    and for the RS (red circles). Galaxies more luminous than \Mr $\approx -23
    \, \mathrm{mag}$ are on average rounder than the galaxies with \Mr $\gtrsim
    -23 \, \mathrm{mag}$. Middle and bottom panel: same as
    \reffig{fig:eps4.vs.logr}, but showing \eps as a function of \Mr for
    $\epsilon < 0.4$ and for $\epsilon \geq 0.4$ . When we
    consider the subsample of flat galaxies ($\epsilon \geq 0.4$), there is no
    trend of \eps with \Mr. This is quantified by the slope of the best-fit
    linear relation, which is statistically consistent with zero (bottom left
    corner of each panel.}\label{fig:eps4.vs.mr}
\end{figure}

In \reffig{fig:eps4.vs.mr} we show \eps for nine bins in \Mr, for the
RC as well as for the reference RS (top panel). For galaxies with
$-22 \, \mathrm{mag} \lesssim \mathrm{M_r} \lesssim -19 \, \mathrm{mag}$, there
seems to be no dependence of \eps with \Mr. However, galaxies more luminous than
$\approx -23 \, \mathrm{mag}$ are on average rounder. In the middle and bottom
panel of \reffig{fig:eps4.vs.mr} we plot \eps against \Mr for the RC
and for the reference RS respectively, but this time we divide each sample
in two two subsets at $\epsilon = 0.4$. Flat galaxies do not present a trend of
\eps with magnitude, and the slope of the best-fit relation is statistically
consistent with zero.

\bgroup
\def\arraystretch{0.5}
\begin{table}
  \small
    \caption[]{The \epsR relation at fixed $r'$-band absolute magnitude (see
    \refsec{subsec:effect.luminosity}).}\label{tab:magtest}
  \begin{tabular}{ccccc}
    \hline
     \Mr & $m_{\rm parent}$ & $\dfrac{m_{\rm parent}}{\sigma_{\rm parent}}$ & $m_{\rm RS}$ & $\dfrac{m_{\rm RS}}{\sigma_{\rm RS}}$ \\
     (1) & (2) & (3) & (4) & (5) \\
    \hline

    $\phantom{-23 < } \mathrm{M_r} \leq -22$ & 0.094 & 2.3 & 0.089 & 2.6 \\
    $-22 < \mathrm{M_r} \leq -21$            & 0.024 & 3.5 & 0.028 & 2.2 \\
    $-21 < \mathrm{M_r} \leq -20$            & 0.038 & 1.5 & 0.058 & 2.1 \\
    $-20 < \mathrm{M_r} \leq -19$            & 0.034 & 2.9 & 0.060 & 4.4 \\
    $-19 < \mathrm{M_r} \leq -18$            & 0.067 & 6.5 & 0.090 & 10.0\\

\hline
\end{tabular}

    Column (1): magnitude interval ($r'$-band absolute magnitude).
    Column (2): the measured slope for the \epsR relation.
    Column (3): significance on the measured slope for the \epsR relation. (in units of the uncertainty $\sigma$).
    Column (4): the measured slope for the \epsR relation for RS galaxies only.
    Column (5): significance on the measured slope for the \epsR relation for RS galaxies only (in units of the uncertainty $\sigma$).

\end{table}
\egroup

Next we studied the \epsR relation by dividing the sample in luminosity bins.
The \epsR relation persists, but decreases in significance
when we decrease the sample size. The results are listed in
\reftab{tab:magtest}. We report the adopted magnitude interval (column 1), the
measured best fit slope (column 2) and the significance of the \epsR relation in
that interval (column 3). We list also the best fit slope for the RS (column 4)
and the significance of the \epsR relation for the RS (column 3). The
significance of the \epsR relation is the ratio of the best fit slope $m$ by its
uncertainty $\sigma$. In general, the \epsR relation is recovered in each
magnitude bin, however the strongest slope and the highest significance are
found for the least luminous galaxies.

\subsection{Galaxy Zoo 2 Morphological Selection}\label{subsec:galaxy.zoo2}

\begin{figure}
   \centering
   \includegraphics[type=pdf,ext=.pdf,read=.pdf,width=0.5\textwidth]{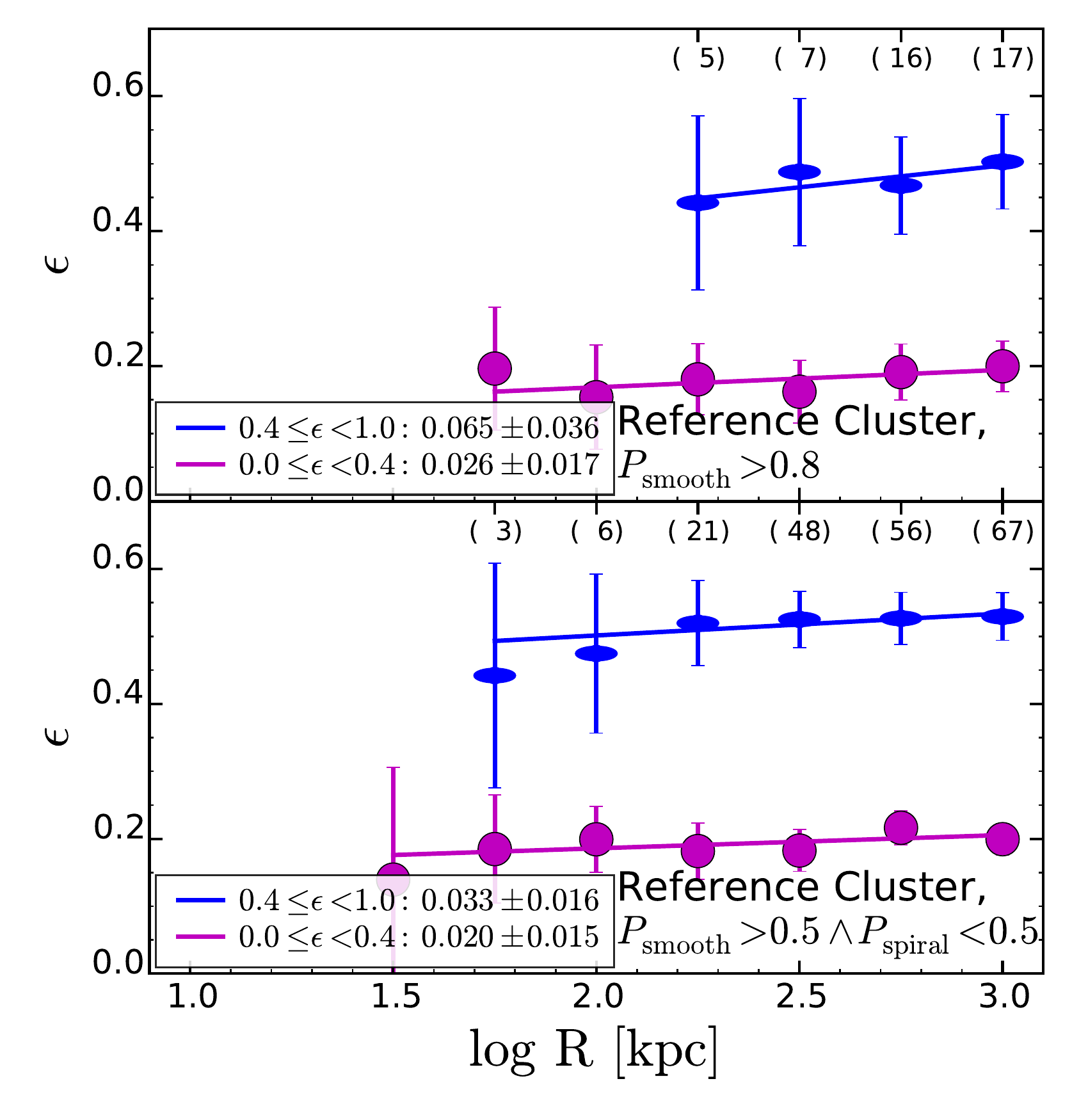}
   \caption{Same as \reffig{fig:eps4.vs.logr}, but for the subsample of cluster
   galaxies morphologically classified as smooth in Galaxy Zoo 2. In each plot,
   the blue ellipses represent galaxies that appear flat on the sky ($\epsilon
   \geq 0.4$). The magenta circles represent galaxies that appear round on the
   sky ($\epsilon < 0.4$). The blue and magenta lines are least squares best
   fits to the data; their slopes are reported in the bottom left corner of each
   figure. The number in parentheses are the number of galaxies in each bin of
   \logR for the flat subsample. The top panel shows the results for the
   subsample of galaxies with $P_\mathrm{smooth} > 0.8$, but it does not have
   enough galaxies to constrain the slope. The bottom panel shows the results
   whith a more generous selection $P_\mathrm{smooth} > 0.5$ and
   $P_\mathrm{spiral} < 0.5$.}\label{fig:eps4.vs.logr.gz2}
\end{figure}

GZ2 is a citizen science project providing
morphological classifications for $\approx 300,000$ galaxies drawn from the SDSS
sample. Cross-correlating their catalogue with our sample we find 1425 matches
($\approx 16 \%$).
We used debiased likelihoods to identify smooth objects ($P_\mathrm{smooth}
\equiv $ \textit{t01\_smooth\_or\_features\_a01\_smooth\_debiased}). Objects with
$P_\mathrm{smooth} > 0.8$ correspond to ETGs in the Hubble classification
\citep{willett2013}. However, this selection leaves us with 252 galaxies, of
which only 50 are flat ($\epsilon \ge 0.4$). This is insufficient to constrain
the slope of the \epsR relation ($m = 0.065 \pm 0.036$, see
\reffig{fig:eps4.vs.logr.gz2}, top panel).

If we relax the above condition and use instead
$P_\mathrm{smooth} > 0.5$ and $P_\mathrm{spiral} < 0.5$, we find 676 galaxies
(and 204 flat galaxies).
In the bottom panel of \reffig{fig:eps4.vs.logr.gz2} we show that the
\epsR relation still holds for flat, smooth galaxies, albeit only to the 2
$\sigma$ level (the best-fit linear slope is $m = 0.033 \pm 0.016$). This is
statistically consistent with the trend without morphological selection ($m =
0.035 \pm 0.006$). Therefore it appears that the spiral contamination is not the
driving mechanism behind the observed relation.

\subsection{Luminosity Profile}\label{subsec:luminosity.profile}

An alternative method to classify galaxies is to use the shape of the luminosity
profile as a function of radius. Following \citet{vincentryden2005}, we used
\textit{fracDeV} to divide the RC in four subsets: galaxies with
$0 <= fracDeV < 0.1 \; (n \lesssim 1.2)$ are ``ex'' galaxies, galaxies with $0.1
<=  fracDeV < 0.5 \; (1.2 \lesssim n \lesssim 2.0)$ are ``ex/de'' galaxies.
Conversely, ``de/ex`` galaxies have $0.5 <= fracDeV < 0.9 \; (2.0 \lesssim n
\lesssim 3.3)$ and finally ``de'' galaxies have $fracDeV >= 0.9 \; (n \gtrsim
3.3)$.

\begin{figure}
   \centering
   \includegraphics[type=pdf,ext=.pdf,read=.pdf,width=0.5\textwidth]{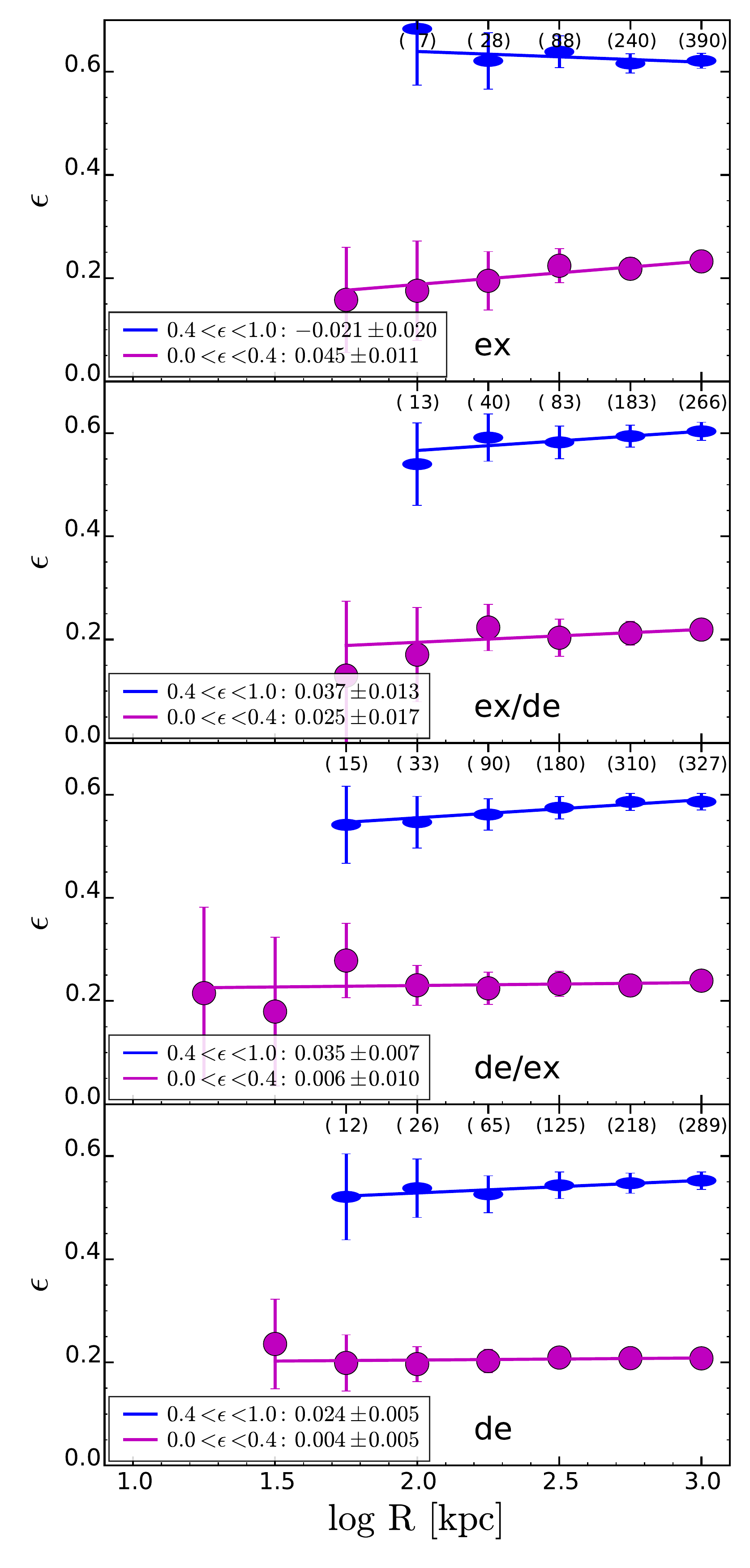}
   \caption{Same as \reffig{fig:eps4.vs.logr}, but here we split the sample
   according to the shape of the luminosity profile. For flat galaxies, the
   trend of increasing \eps with \logR is recovered for de/ex and de galaxies
   (third and fourth panel). For flat ex/de galaxies there is only marginal
   evidence of a trend (second panel), while no significant trend is observed for flat ex
   galaxies (top panel). See \refsec{subsec:luminosity.profile} for the
   definition of the labels.}\label{fig:eps4.vs.logr.kr05}
\end{figure}

In \reffig{fig:eps4.vs.logr.kr05} we plot the \epsR relation for the RC,
splitting the galaxies according to their light profile. For flat ex
galaxies the best-fit slope shows that there is no evidence of correlation
between \eps and \logR (top panel). For flat ex/de galaxies there is marginal
evidence for a trend (the slope of the best-fit linear relation is 2 $\sigma$
above 0, second panel). For flat de/ex and de galaxies the significance
increases to 5 $\sigma$ (third and bottom panels).
The fraction of flat ex and ex/de galaxies increases with \logR (numbers in
parentheses). This trend, combined with the fact that ex and ex/de galaxies
are on average a little flatter than de/ex and de galaxies, contributes to the
\epsR relation observed for the undivided RC (see
\refsec{subsec:full.sample}).\\

\section{Discussion}\label{sec:discussion}

We have shown that in rich clusters \eps depends on radius. We find that
galaxies closer to the cluster centre are on average rounder. The \epsR
relation persists when considering only galaxies with $\epsilon \geq 0.4$.
In the next section we evaluate the effect of observational bias
and establish that the trend is indeed genuine. We then evaluate the effect of
the luminosity (\refsec{subsec:effect.luminosity}) and of the \TS relation
(\refsec{subsec:morphology.density.relations}). Finally we discuss two related
works (\refsec{subsec:other.works}) and proceed to review the possible
explanations for the observed relation (\refsec{subsec:possible.driving.mechanisms}).

\subsection{Observation bias}\label{subsec:observation.bias}

The value of \eps is already corrected for PSF effects, because the model for
the light profile is convolved with the relevant PSF. To evaluate any residual
PSF effect we conduct two tests. We repeated our analysis
using $\epsilon_\mathrm{moments}$ instead of $\epsilon_\mathrm{model}$,
(as in e.g. KR05). $\epsilon_\mathrm{moments}$ was corrected
for PSF effects according to the prescription of \citet{bernsteinjarvis2002}.
Given that $\epsilon_\mathrm{model}$ is systematically higher than
$\epsilon_\mathrm{moments}$, it is not surprising that our result
changes. However the observed trends are still present (see
Appendix \ref{app:moments.ellipticity}). As a further test for the
effect of the PSF, we repeated our analysis limiting our sample to well resolved
objects (where $\tau / \tau_{\mathrm{PSF}} > 6.25$, see KR05).
The \epsR relation persists even for these galaxies where the effect of the PSF
is negligible, with lower significance as expected from the smaller sample size.\\

Finally, we remark that the PSF would make all galaxies
appear rounder, affecting smaller galaxies more than larger ones. This means
that in order for PSF effects to cause or enhance the \epsR relation, smaller
galaxies would need to be a larger fraction of the population at small $R$, but
the opposite is true \citep{dressler1980, hogg2004}.\\

All of our samples include a number of interlopers, foreground or background
galaxies which are not gravitationally bound to the relevant cluster. This is
true both for the parent sample and (to a lesser extent) for the RS sample.
To understand the possible effect of interlopers on the observed trend of
\eps versus \logR, we consider the following scenario. For a given cluster,
we assume that interlopers are distributed uniformly across the sky. In
contrast, the density of cluster member galaxies is higher in the centre and
decreases at larger radii. This means that the fraction $f$ of interlopers over
the sample increases with \logR. If interlopers had a higher average \eps than
cluster members, then the observed variation of \eps with \logR would be a
consequence of the variation of $f$ with \logR.  For \Ab 1656 this scenario is
ruled out by the fact that the \epsR relation holds even for the sample of
spectroscopically confirmed cluster members
(\refsec{subsec:spectroscopic.a1656}). In addition, the 61 interlopers have an
average ellipticity $\epsilon = 0.34 \pm 0.19$, lower than the value for cluster
members. As expected, the fraction of interlopers increases with \logR.
If interlopers have a lower than average \eps for the other clusters too, then their
inclusion cannot artificially create the observed \epsR relation, but rather acts to mask it.

\subsection{The effect of luminosity}\label{subsec:effect.luminosity}

More luminous galaxies are both rounder and more common in the
densest environments \citep{dressler1980, hogg2004}. Could the \epsR
relation arise from this trend? This hypothesis is contradicted
by the fact that for the flat subsample there is no trend of \eps with \Mr
(\reffig{fig:eps4.vs.mr}), but for the same galaxies a clear correlation exists
between \eps and \logR. In addition, the trend of \eps with \logR is observed
even at fixed magnitude, and is stronger for the less luminous galaxies
(\reftab{tab:magtest}).

\subsection{The morphology-density relations}\label{subsec:morphology.density.relations}

The final source of bias that we consider is the \TS relation \citep{dressler1980}.
Es, S0s and spiral galaxies have different \eps distributions. Therefore the trend
of their number fraction with local environment affects systematically the
overall \eps distribution. In order to quantify this effect, we need Hubble
morphological classifications for all the galaxies, but these are not available.

In \refsec{subsec:flat.galaxies} we divided the RC (and the
reference RS) into a flat ($\epsilon \ge 0.4$) and a round subsample ($\epsilon
< 0.4$). The subset of round galaxies contains both SRs (regardless of their
inclination on the sky) and close-to-face-on discs. Therefore the observed
trend of increasing flattening with \logR can be due to the decreasing fraction
of SRs \citep{cappellari2011b, deugenio2013, houghton2013, scott2014, fogarty2014}
and/or to a change in the intrinsic shape of galaxies. Without a dynamical
classification for each galaxy it is impossible to draw any conclusions from the
observed trend.

However the subset of galaxies with $\epsilon \geq 0.4$ does not contain SRs
(which are intrinsically round). Therefore the observed trend of increasing
projected ellipticity with radius arises because flat galaxies have lower
intrinsic ellipticity nearer the centre of clusters. Flat galaxies include both
FR ETGs and spiral galaxies. If these two classes have different \distr, the
\epsR relation could result from a trend of the morphological fractions with
cluster-centric radius. However, it seems that on average FR ETGs have the same
intrinsic flattening as spiral galaxies \citep{weijmans2014}, a fact that
undermines the above explanation. Nevertheless in \refsec{subsec:galaxy.zoo2} we
attempted to remove spiral galaxies: we used GZ2 data to select a sample of smooth
objects. When we adopt the condition $P_\mathrm{smooth} > 0.8$, we find 252
galaxies, of which only 50 are flat ($\epsilon \ge 0.4$). The slope of \epsR
relation for flat galaxies is 1.8 standard deviations above 0
(\reffig{fig:eps4.vs.logr.gz2}, top panel). For a sample of 50
flat galaxies selected randomly from the RC, we expect the slope
to be 0.6 standard deviations away from 0.
This suggests (but does not demonstrate) that the low significance of the \epsR
relation of smooth, flat galaxies is due to insufficient sample size, as opposed
to the systematic removal of the galaxies with $P_\mathrm{smooth} \le 0.2$. We
then relaxed the morphological selection, finding 676 galaxies, of which 204 are
flat (we used $P_\mathrm{smooth} > 0.5$ and $P_\mathrm{spiral} < 0.5$, see again
\refsec{subsec:galaxy.zoo2}). In this case the slope of the \epsR relation of
flat galaxies was 2.1 standard deviations above 0.
For a sample of 204 flat galaxies randomly selected from the RC,
we expected the slope to be 1.3 standard deviations above 0. The fact that we
find a value of 2.1 suggests again that the low significance is due to the
sample size rather than to any systematic effect. In addition, the slope is
statistically consistent with what found in \reffig{fig:eps4.vs.logr}, even
though a significant number of spiral galaxies have been removed.

\citet{willett2013} cross-correlated the GZ2 catalogue with a professionally
classified sample of SDSS galaxies \citep{nairabrham2010}. They showed that galaxies
classified as smooth in GZ2 ($P_\mathrm{smooth} > 0.8$) are morphological ETGs
for 96.7 \%  of the cases (the matching for our relaxed conditions is not
covered). Therefore it appears that the spiral contamination is not the driving
mechanism behind the observed relation.\\

Our result could still be explained by the \TS relation, because
within ETGs the fraction of Es increases with projected density \citep[and nearer
to the centre of clusters, ][]{whitmore1993}. Since Es are on average rounder
than S0s, the \epsR relation could be an effect of the \TS relation. To
investigate this possibility we need to distinguish between Es and S0s, which
cannot be done with GZ2 and is beyond the scope of this paper. But the latest
view on the E/S0 divide, offered by the \atl survey, is that the traditional
visual E/S0 morphology is flawed.
\citet{emsellem2011} demonstrated that two thirds
of the locally classified Es are kinematically identical to S0s. We therefore
reframe our findings within the FR/SR paradigm. 15\% of the local ETGs are SRs,
but their fraction increases dramatically nearer the centre of clusters.
\citep{cappellari2011b, deugenio2013, houghton2013, scott2014, fogarty2014}.
Since
SRs are on average rounder than FRs, the \epsR relation could be a consequence
of the kinematic morphology-density relation. 
However, by selecting galaxies with $\epsilon \geq 0.4$, we eliminate all SRs.
By further imposing $P_\mathrm{smooth} > 0.8$ we eliminate all spiral galaxies.
The fact that the \epsR relation persists even for smooth galaxies flatter than
$\epsilon = 0.4$ demonstrates that the \epsR relation is not a consequence of the kinematic
morphology-density relation. The decrease of projected ellipticity \eps nearer
the centre of clusters implies that the distribution of \textit{intrinsic}
ellipticity changes with \logR. This is evidence of the effect of the cluster and/or local
environment on the shape of FRs, be they FR Es or FR S0s.\\

\subsection{Comparison with other work}\label{subsec:other.works}

KR05 used a much larger sample of $\approx 200,000$ galaxies to investigate the
relation between the shape of galaxies (as expressed by the axis ratio $q$) and
the local environment (measured by the number density of neighbour galaxies in
the sky). Their sample differs from ours in that: (i) they adopted a cut in
apparent magnitude at $r' = 17.77 \; \mathrm{mag}$ and (ii) their selection is
not limited to specific environments, whereas we focussed on rich clusters of
galaxies. These differences make it difficult to compare our results.

The noise introduced by their selection can be appreciated by looking at their
Tables 3-5. Despite having $\approx 20 \times$ the sample size, the significance
of the recovered trends is similar to ours.
KR05 divide their sample according to the luminosity profile, as we did in
\refsec{subsec:luminosity.profile}. Our results are in qualitative agreement
with theirs. However they do not investigate the trend of intrinsically flat
galaxies.\\

\citet{weijmans2014} studied the intrinsic shape distribution of the \atl ETGs,
using Integral Field Spectroscopy to separate SRs and FRs. As we have seen, they
find that the intrinsic flattening of FR ETGs is consistent with that of spiral
galaxies, a fact that reinforces our result. However, contrary to our findings,
they do not observe any correlation of intrinsic flattening and environment (as
parametrised by \logsig3). It is unclear if this depends on the small size of
their sample, and/or on the range of environments that they explore (the \atl
survey contains just one unrelaxed cluster, Virgo).

\subsection{Possible driving mechanisms}\label{subsec:possible.driving.mechanisms}

Here we propose three possible explanations for the observed trend.
\begin{itemize}
  \item[(a)] We can view our result in the framework of the B/D decomposition.
  By definition discs are intrinsically flat, while bulges are rounder.
  A systematic change in the B/D mass ratio with \logR could therefore account for
  the \epsR relation. We notice however that in their study of eight nearby clusters
  ($z < 0.06$), \citet{hudson2010} do not find any correlation between R-band B/D
  light ratios and the cluster-centric radius (except for the innermost $\approx 0.3$ Mpc).
  \item[(b)] The cluster environment affects the stellar population of member
  galaxies, for example by removing cold gas from the disc and halting star
  formation there \citep[see][ for a review]{boselligavazzi2006}. We therefore
  expect to observe a trend between the cluster-centric radius and the photometric
  properties of galaxies. For instance, even at
  fixed B/D mass ratio, a trend in stellar age can generate a trend in
  the B/D light ratio, which in turn could give rise to the \epsR relation. This
  is supported by the observed correlation between disc colour and cluster
  centric radius, while no such correlation is observed for bulge colour
  \citep{balogh2000, hudson2010, head2014}.
  \item[(c)] FRs might contain subclasses with different intrinsic ellipticity,
  for instance \citet{weijmans2014} suggests the existence of a tail of rounder
  FRs in the \atl sample. If this is correct, the \epsR relation for FRs could
  arise from the radial distribution of different classes of FRs.
  \item[(d)] Cluster environments also affect the dynamical structure of
  galaxies. Harassment and stripping make discs smaller and thicker, therefore
  changing their intrinsic ellipticity. This is supported by
  \citet{houghton2013}, who studied a sample of galaxies in the central 15$'$
  (radius) of the Coma Cluster. For each galaxy they measure $\lambda$ \citep[a
  proxy for the projected angular momentum
  per unit mass][]{emsellem2007}, and find that the maximum value of $\lambda$
  for Coma FRs is lower than that of FRs from the \atl survey
  \citep{emsellem2011}. They suggest
  that this is due to the cluster environment affecting the anisotropy
  \citep{binneytremaine1988} of galaxies.
\end{itemize}

In order to understand the origin of the \epsR relation, an extended integral
field spectroscopic survey of cluster galaxies is required. The Coma Cluster
is an ideal starting point. Ongoing surveys \citep[SAMI, MaNGA; ][]{croom2012, bundy2015}
could also address this problem, depending on their coverage of rich
clusters.

\section{Conclusion}\label{sec:conclusion}

We studied the projected ellipticity \eps of galaxies in a
sample of twenty nearby ($z < 0.1$) rich clusters and show that:
\begin{itemize}
  \item[(i)] there is no evidence of differences between the distribution of \eps
  among the clusters in our sample, except possibly for the Coma Cluster (\Ab 1656)
  \item[(ii)] there exists a correlation between \eps and the projected cluster-centric
  radius $R$, which we refer to as the \epsR relation
  \item[(iii)] the \epsR relation exists independently of the trend of more luminous galaxies
  to be both rounder and more common in the centre of clusters
  \item[(iv)] the \epsR relation is stronger for galaxies on the RS, and
  persists for a redshift selected sample in the Coma Cluster
  \item[(v)] the \epsR relation is steeper for flat galaxies ($\epsilon \geq
  0.4$; excludes SRs) than for round galaxies ($\epsilon < 0.4$; a
  mixture of FRs and SRs)
  \item[(vi)] for flat galaxies, there is no correlation between \eps and 
  $r'$-band luminosity
  \item[(vii)] there is marginal evidence (2 $\sigma$) that the \epsR relation persists
  for flat ETGs, as classified by GZ2. The low significance is likely due to the small
  sample size.
\end{itemize}
We concluded that the \epsR is evidence for physical effects causing intrinsically flat
galaxies (including FR ETGs) to be rounder near the centre of rich clusters.

\section*{Acknowledgments}

We acknowledge the referee Eric Emsellem for many constructive comments.

This work was supported by the Astrophysics at Oxford grants (ST/H002456/1 and
ST/K00106X/1) as well as visitors grant (ST/H504862/1) from the UK Science and
Technology Facilities Council.
RLD acknowledges travel and computer grants from Christ Church, Oxford, and support
for a visit from the Space Telescope Science Institute in the final stages of
this work.

RCWH was supported by the Science and Technology Facilities Council [STFC
grant numbers ST/H002456/1 \& ST/K00106X/1].

FDE acknowledges studentship support from the Department of Physics,
Oxford University, and the Simms Bursary from Merton College,
Oxford.

EDB was supported by Padua University through grants 60A02-4807/12, 60A02-5857/13, 60A02-5833/14, and CPDA133894. EDB acknowledges the Sub-department of Astrophysics, Department of Physics, University of Oxford and Christ Church College for their hospitality while this paper was in progress.

Funding for SDSS-III has been provided by the Alfred P. Sloan Foundation, the
Participating Institutions, the National Science Foundation, and the U.S.
Department of Energy Office of Science. The SDSS-III web site is
http://www.sdss3.org/.

Finally, we acknowledge the work of the Galaxy Zoo 2 team and volunteers.

\selectlanguage{english}

\bibliographystyle{mn2e}

\appendix
\section{Moments ellipticity}\label{app:moments.ellipticity}

KR05 use the weighted second order moments of the $r'$-band
surface brightness $I$ to measure the shape of galaxies. Following
\citet{bernsteinjarvis2002} they use a Gaussian weight function $w(r, c)$ matched
to the size and shape of the galaxy being measured ($r$ and $c$ are the rows and
columns in the image). If we define:
\begin{equation}
    \langle f \rangle_w \equiv \dfrac{\int f(r, c) w(r, c) I(r, c) dr \, dc}{\int w(r, c) I(r, c) dr \, dc}
\end{equation}
then the centre of the galaxy is given by:
\begin{equation}
    (r_0, c_0) \equiv (\langle r \rangle_w, \langle c \rangle_w) 
\end{equation}
and the (central) second moments are defined as:
\begin{align}
\begin{split}
  M_{rr} &\equiv \langle (r - r_0)^2 \rangle_w \\
  M_{rc} &\equiv \langle (r - r_0) (c - c_0) \rangle_w \\
  M_{cc} &\equiv \langle (c - c_0)^2 \rangle_w
\end{split}
\end{align}
From these quantities we can define:
\begin{align}
\begin{split}
  \tau & \equiv M_{cc} + M_{rr} \rightarrow \text{mRrCr} \\
  e_+ & \equiv \dfrac{M_{cc} - M_{rr}}{\tau} \rightarrow \text{mE1} \\
  e_\times & \equiv \dfrac{2 \, M_{rc}}{\tau} \rightarrow \text{mE2} \\
  e & \equiv \sqrt{e_+^2 + e_\times^2}
\end{split}
\end{align}
where \textit{mRrCc}, \textit{mE1} and \textit{mE2} are the corresponding
entries in the SDSS database. The axis ratio is given by:
\begin{equation}
  q \equiv \sqrt{\dfrac{1 - e}{1 + e}}
\end{equation}
and finally we define the projected ellipticity $\epsilon$ as:
\begin{equation}
    \epsilon \equiv 1 - q
\end{equation}
In order to correct for the effect of the PSF, we follow again
\citet{bernsteinjarvis2002}, which use the fourth order moment as well as the
moments of the PSF itself (this data is included in the SDSS database).\\

\begin{figure}
   \centering
   \includegraphics[type=pdf,ext=.pdf,read=.pdf,width=0.5\textwidth]{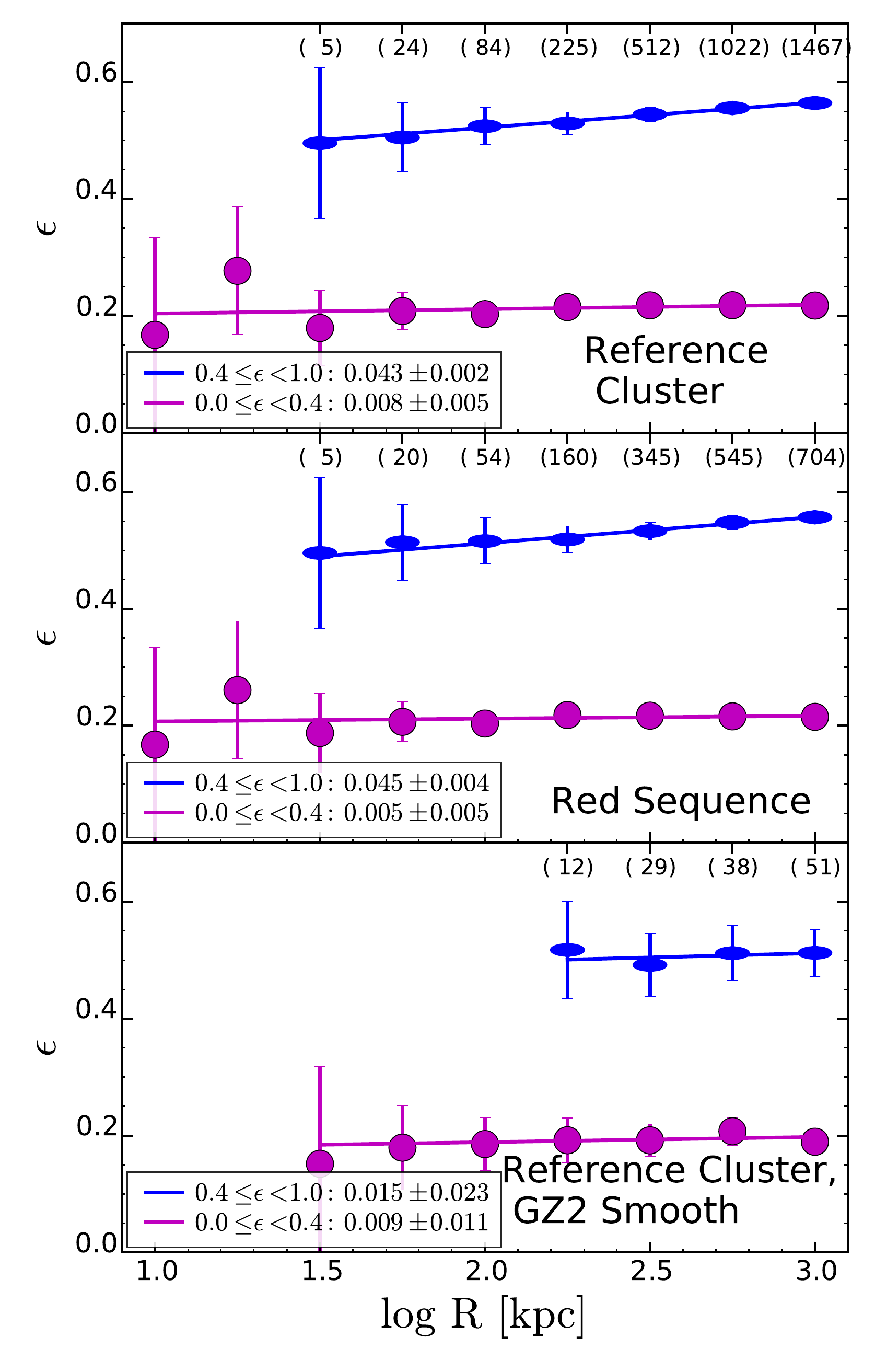}
   \caption{Same as \reffig{fig:eps4.vs.logr}, but using the projected moments
   ellipticity $\epsilon_\mathrm{moments}$. The method of moments, as implemented
   in SDSS, gives smaller values than $\epsilon_\mathrm{model}$. The sample of
   flat galaxies ($\epsilon \geq 0.4$) is therefore smaller, and it is impossible
   to assess any trend from the GZ2 cross-correlated sample.}\label{fig:eps4mom.vs.logr}
\end{figure}

When we use $\epsilon \equiv \epsilon_\mathrm{moments}$ the \epsR relation is
still observed (\reffig{fig:eps4mom.vs.logr}). However, given that generally we
have $\epsilon_\mathrm{moments} < \epsilon_\mathrm{model}$
(\refsec{subsec:projected.ellipticity}), we have
fewer flat galaxies ($\epsilon_\mathrm{moments} \geq 0.4$). Therefore we cannot
use $\epsilon_\mathrm{moments}$ to constrain the \epsR relation of smooth, flat
galaxies (bottom panel in \reffig{fig:eps4mom.vs.logr}).

\label{lastpage}

\end{document}